\newif\ifconfver
\confvertrue        

\newif\ifplainver  
\plainvertrue

\newif\ifhide  
\hidetrue

\ifplainver
    \confverfalse   
\fi

\ifconfver
     \documentclass[10pt,twocolumn,twoside]{IEEEtran}
\else
    \ifplainver
        \documentclass[11pt]{article}
        \usepackage{fullpage}
    \else
        \documentclass[12pt,draftcls,onecolumn]{IEEEtran}
    \fi
\fi

\usepackage{calc,amsfonts,amssymb,amsmath,bm,url,color,theorem,graphicx,cite}
\usepackage{psfrag,float}
\usepackage{algorithm}
\usepackage{algorithmic}
\usepackage{soul}
\usepackage{enumerate}
\usepackage{bbm}
\usepackage{multirow}
\usepackage{subcaption}
\usepackage{array}
\usepackage{epstopdf}

\usepackage{etoolbox}%
\usepackage{xpatch}
\usepackage{blindtext}
\usepackage{tocloft}%
\newlength{\articlesectionshift}%
\let\LaTeXStandardSection\section
\let\LaTeXStandardTheSection\thesection
\let\LaTeXStandardTheSubSection\thesubsection
\let\LaTeXStandardTheSubSubSection\thesubsubsection
\let\LaTeXStandardTheParagraph\theparagraph

\makeatletter
\newcounter{titlecounter}

\xpretocmd{\maketitle}{\ifnumgreater{\value{titlecounter}}{1}}{\clearpage}{}{} 
\xpatchcmd{\maketitle}{\let\maketitle\relax\let\@maketitle\relax}{\refstepcounter{titlecounter}\begingroup
	\addtocontents{toc}{\begingroup\addtolength{\cftsecindent}{-\articlesectionshift}}%
	\addcontentsline{toc}{section}{\protect{\numberline{\thetitlecounter}{\@title~ \@author}}}%
	\addtocontents{toc}{\endgroup}
}{%
	\typeout{Patching was successful}
}{%
	\typeout{patching failed}
}%

\def\@IEEEdestroythesectionargument#1{\LaTeXStandardSection{#1}}%

\xapptocmd{\maketitle}{%
	\renewcommand{\thesection}{\LaTeXStandardTheSection}%
	\renewcommand{\thesubsection}{\LaTeXStandardTheSubSection}%
	\renewcommand{\thesubsubsection}{\LaTeXStandardTheSubSubSection}%
	\renewcommand{\theparagraph}{\LaTeXStandardTheParagraph}%
}{}{}%

\@addtoreset{section}{titlecounter}

\newlength{\customfigwidth}
\newcolumntype{M}[1]{>{\centering\arraybackslash}m{#1}}

\definecolor{orange}{RGB}{255,107,0}

\newtheorem{Fact}{Fact}

\newtheorem{Theorem}{Theorem}

\theorembodyfont{\rmfamily}

\newtheorem{Remark}{Remark}

\newcommand\bw{\ensuremath{{\bm w}}}
\newcommand\bW{\ensuremath{{\bm W}}}

\newcommand\bx{\ensuremath{{\bm x}}}
\newcommand\by{\ensuremath{{\bm y}}}
\newcommand\bG{\ensuremath{{\bm G}}}
\newcommand\bh{\ensuremath{{\bm h}}}
\newcommand\bH{\ensuremath{{\bm H}}}

\newcommand\bz{\ensuremath{{\bm z}}}

\newcommand\bR{\ensuremath{{\bm R}}}

\newcommand\bb{\ensuremath{{\bm b}}}
\newcommand\bg{\ensuremath{{\bm g}}}

\newcommand\bu{\ensuremath{{\bm u}}}
\newcommand\bv{\ensuremath{{\bm v}}}

\newcommand\btheta{\ensuremath{{\bm \theta}}}

\newcommand{\Rbb}{\mathbb{R}}
\newcommand{\Cbb}{\mathbb{C}}

\newcommand{\setX}{\mathcal{X}}

\newcommand{\setD}{\mathcal{D}}

\newcommand{\Exp}{\mathbb{E}}

\newcommand{\Rfrak}{\mathfrak{R}}
\newcommand{\Ifrak}{\mathfrak{I}}
\newcommand\jj{\ensuremath{{\frak j}}}

\newcommand{\bI}{{\bm I}}

\newcommand\sgn{\ensuremath{{\rm sgn}}}

\newcommand{\beq}{\begin{equation}}
\newcommand{\eeq}{\end{equation}}

\begin{document}

\bibliographystyle{IEEEtran}

\newcommand{\papertitle}{
Binary MIMO Detection via Homotopy Optimization and Its
Deep Adaptation
}

\newcommand{\paperabstract}{
In this paper we consider maximum-likelihood (ML) MIMO detection under one-bit quantized observations and binary symbol constellations.
This problem is motivated by the recent interest in adopting coarse quantization in massive MIMO systems---as an effective way to scale down the hardware complexity and energy consumption.
Classical MIMO detection techniques consider unquantized observations,
and many of them are not applicable to the one-bit MIMO case.
We develop a new non-convex optimization algorithm for the one-bit ML MIMO detection problem, using a strategy called homotopy optimization.
The idea is to transform the ML problem into a sequence of approximate problems,
from easy (convex) to hard (close to ML),
and with each problem being a gradual modification of its previous.
Then, our attempt is to iteratively trace the solution path of these approximate problems.
This homotopy algorithm is well suited to the application of deep unfolding, 
a recently popular approach for turning certain model-based algorithms into data-driven, and performance enhanced, ones.
While our initial focus is on one-bit MIMO detection,
the proposed technique also
applies naturally to the classical unquantized MIMO detection.
We performed extensive simulations and show that the proposed homotopy algorithms, both non-deep and deep, have satisfactory bit-error probability performance compared to many state-of-the-art algorithms.
Also, the deep homotopy algorithm has attractively low computational complexity.
}


\ifplainver


    \title{\papertitle}

    \author{
    Mingjie Shao and Wing-Kin Ma \\
    $^\dag$Department of Electronic Engineering, The Chinese University of Hong Kong, \\
    Hong Kong SAR of China
    }

    \maketitle

    \begin{abstract}
    \paperabstract
    \end{abstract}

\else
    \title{\papertitle}

    \ifconfver \else {\linespread{1.1} \rm \fi

    \author{Mingjie Shao and Wing-Kin Ma
    }

    \maketitle

    \ifconfver \else
        \begin{center} \vspace*{-2\baselineskip}
        \end{center}
    \fi

    \begin{abstract}
    \paperabstract
    \end{abstract}


    \begin{IEEEkeywords}\vspace{-0.0cm}
        MIMO detection, one-bit MIMO, homotopy optimization, deep unfolding
    \end{IEEEkeywords}

    \ifconfver \else \IEEEpeerreviewmaketitle} \fi

 \fi

\ifconfver \else
    \ifplainver \else
        \newpage
\fi \fi

\section{Introduction}

\ifconfver
{\let\thefootnote\relax\footnotetext{This work was supported by a General Research Fund (GRF) of the Research Grant Council (RGC), Hong Kong, under Project ID CUHK	142017318.}}
\fi

MIMO detection has been, for decades, at the center of some very important developments in signal processing, communications and beyond.
Its problem statement is simple---to detect a multitude of  finite-alphabet variables from a noisy observation of their linearly entangled counterparts.
The problem is, however, very challenging if
the goal is
to build an MIMO detection algorithm that can achieve optimal or near-optimal detection performance with a low implementation cost, particularly, for large problem sizes.
The above goal
has attracted generations of researchers' innovations, e.g., around the 1990's for multiuser code division multiple access, around the 2000's for multi-antenna systems, and recently for massive antenna arrays;
and the consequence is a rich, enduring, collection of  methods---such as sphere decoding methods 
\cite{damen2003maximum,jalden2005complexity}
for implementing the optimal maximum-likelihood (ML) detection via combinatorial search;
convex relaxation methods \cite{tan2001constrained,yener2002cdma,tan2001application,ma2002quasi,thrampoulidis2018symbol} for efficient ML approximation via convex optimization;
lattice reduction-aided methods 
\cite{wubben2011lattice}
for improving the channel conditioning, and thereby performance, via basis design under integer coefficient constraints;
soft detection methods 
\cite{luo2001near,fertl2011performance}
for posterior inference of the unknown finite-alphabet variables via various approximations;
to name a few.
These developments show intimate connections with studies in other areas, such as complexity theory, combinatorial and continuous optimization, statistical inference, etc.
Also,
recent research takes insight from approximate message passing in compressive sensing
\cite{maleki2013asymptotic}
to tackle MIMO detection \cite{jeon2018optimal}.
Hence, MIMO detection is a topic of far-reaching implications.
We refer the reader to the literature, e.g., \cite{yang2015fifty}, for a chronological account  of the MIMO detection developments and a coverage of the numerous algorithms therein.

Lately, MIMO detection has been igniting new interest in two recent research directions, namely, coarsely quantized massive MIMO and deep learning for communications.
We separately describe them in the following two subsections.

\subsection{Coarsely Quantized Massive MIMO Detection}

As a serious impediment to the real-world massive MIMO implementation,
it is well-known that if every radio-frequency front end of a massive MIMO system is equipped with a high-resolution analog-to-digital converter (ADC) or digital-to-analog converter (DAC)---as in conventional MIMO, the system will be very expensive to implement in terms of hardware complexity and energy consumption.
In view of this, there has been movement toward the use of low-resolution ADCs or DACs in massive MIMO.
One arising problem in this context is MIMO detection
under coarsely quantized observations---particularly one-bit, or sign-only, observations.

A challenge with coarsely quantized MIMO detection is that the classical MIMO detection methods we have studied for decades cannot be straightforwardly extended to the quantized case;
this is owing to the different likelihood function in the quantized case.
Some powerful classical methods, such as sphere decoding, semidefinite relaxation \cite{tan2001application,ma2002quasi} and lattice reduction, are inapplicable in the quantized case because they were designed to exploit the convex quadratic structure of the ML objective function in the unquantized case.
Some researchers study the impacts of coarse quantizations on some classical MIMO detection methods (e.g., zero forcing)  \cite{risi2014massive},
while others redesign the MIMO detection methods to better harness the problem structure.
For the latter,
Studer and Durisi took inspiration from the classical linear minimum-mean-square-error (MMSE) detection to derive a variant in the quantized case \cite{studer2016quantized};
Choi, Mo and Heath developed a convex sphere relaxation of one-bit ML MIMO detection \cite{choi2016near};
Jeon {\em et al.} devised a sphere decoder for one-bit ML MIMO detection~\cite{jeon2018one} (its appearance is very different from those we see in classical MIMO detection);
approximate message passing was redesigned for quantized MIMO detection in
\cite{wang2014multiuser,wen2015bayes};
coding theory-inspired detection algorithms were introduced in \cite{hong2017weighted,cho2019one,kim2019supervised}.
Also, quantized MIMO detection for orthogonal frequency-division modulation (OFDM)-MIMO---a more realistic, but also computationally more challenging, scenario---was studied in \cite{studer2016quantized,plabst2018efficient,Mirfarshbafan2020,shao2020divide}.


\subsection{Deep Learning for MIMO Detection}

Recently, the tremendous successes of deep neural networks in natural language processing, computer vision and machine learning have sparked widespread interest in the communications community.
In particular, we have seen a variety of emerging deep network applications for communications, such as
end-to-end communication system design
\cite{o2017introduction}, 
multiuser power control \cite{sun2018learning},
and, as our main interest, MIMO detection
\cite{samuel2017deep,Samuel2019learning,takabe2019trainable,corlay2018multilevel,Tan2020improving,He2020model,mohammadkarimi2019deep}.
Neural networks for MIMO detection
was already considered in around the 1990's \cite{gibson1989multilayer,chen1993clustering};
it is worth noting that
the motivation at the time lies in neural networks' ability to generate nonlinear decision regions, which the then-popular methods of linear and decision-feedback detection have limitations.
In the recent renewed interest, we have  so far not seen
a report on
successfully
training a deep neural network---specifically,
that under a standard network architecture---that gives consistently near-optimal detection performance and good training
stability
for a broad range of MIMO settings.
Instead, attention appears to have been drawn to
the design approach of
{\em deep unfolding}, which
has a flavor of leveraging
on both the existing model-based MIMO detection methods and the data-driven deep learning approach.

Deep unfolding was first introduced in the context of sparse coding by Gregor and LeCun  \cite{gregor2010learning},
and it has recently received much attention
\cite{sprechmann2015learning,chen2018theoretical}. 
The rationale is to see an existing iterative algorithm, e.g., the
iterative shrinkage thresholding algorithm (ISTA)
in sparse coding (see \cite{gregor2010learning}), as a deep network.
It intends to learn a better algorithm than its predecessor algorithm by untying some parameters of the predecessor, and then by learning those parameters from data.
Also, one can modify part of the structure of the predecessor to make it more general, and learn the new structure
from data.
The result is a structured deep network whose structures preserve some of the structures prescribed under a model-based framework.
In the work by Gregor and LeCun, they showed that the learnt ISTA requires much less number of iterations, or layers, to achieve performance comparable to ISTA.

Deep unfolding was first applied to MIMO detection by Samuel, Diskin and Wiesel in 2017 \cite{samuel2017deep}.
The algorithm there, called DetNet, is a deep unfolding of a non-convex projected gradient algorithm for ML MIMO detection.
We have seen growing interest with deep learning for MIMO detection since the DetNet work.
For example, the works \cite{takabe2019trainable,corlay2018multilevel}
studied deep network structures similar to DetNet.
The works \cite{Tan2020improving,He2020model} studied the deep unfolding of approximate message passing.
Apart from deep unfolding,
the work \cite{mohammadkarimi2019deep} applied deep learning to learn the sphere radius in sphere decoding.


\subsection{Present Contribution and Related Works}

The present contribution considers one-bit ML MIMO detection under binary symbol constellations.
We tackle the problem by a non-convex optimization strategy, namely, homotopy optimization; see
\cite{wu1996effective,dunlavy2005homotopy,xiao2013proximal,mobahi2015link,hazan2016graduated,anandkumar2017homotopy} and the references therein.
Also called continuation or graduated non-convexity, homotopy optimization is an idea that arose independently from a variety of applications in different fields, such as  molecular conformation \cite{wu1996effective}, sparse optimization \cite{xiao2013proximal} and neural network training \cite{hazan2016graduated}.
Homotopy methods can be vastly different from one application to another,
but they often follow a common principle.
Specifically, it entails a problem transformation---called homotopy map---that has the flexibility of either making the transformed problem easier to solve (e.g., convex) but less accurate in approximating the original problem; or making the transformed problem harder but closer to the original.
Then, the attempt is to iteratively trace the solution path of a sequence of such approximate problems, from easy to hard and in a gradual fashion.
In the context of MIMO detection, our empirical experience is the following:
tackling the ML MIMO detection problem directly by a straight application of a non-convex algorithm may result in poor detection performance, owing to  convergence to poor local minima;
but handling the problem indirectly via the aforementioned gradually easy-to-hard optimization principle may lead to better performance.

Homotopy optimization offers us a principle, not a numerical algorithm that fits all problems.
We often need to find a suitable problem transformation for the application at hand.
We
employ the non-convex continuous reformulation of binary optimization in \cite{shao2019framework}, as well as the efficient first-order algorithm design therein.
This technique was previously proposed by us to tackle a one-bit MIMO precoding problem.
In the present contribution, we incorporate this technique into the theme of homotopy optimization and explore its potential in MIMO detection.
Also we enrich the result by connecting homotopy optimization with Lagrangian dual relaxation (LDR), as will be shown in Section~\ref{sect:homotopy_ldr}.
It is worth noting that the LDR notion plays a vital role in ML MIMO detection; e.g., semidefinite relaxation and regularized lattice decoding can be interpreted as outcomes of LDR
\cite{poljak1995recipe,pan2013mimo}.

In our development, we found that the homotopy algorithm has a  structure favorable for deep unfolding.
This motivates to consider deep unfolding of our homotopy algorithm.
Our deep unfolding involves mild untying and structure modification, which means that the structure does not change a lot.
By empirical experience, our deep homotopy network is easy to train.
Also our deep homotopy network is trained to cater for different channels, rather than a fixed channel.
Note that the former is more preferable since it allows us to have the expensive training done offline;
the latter requires online training for every given channel, and the subsequent real-time computational overheads may be significant.


The contributions of this work are summarized as follows.
\begin{enumerate}[1.]
	\item As a new attempt to tackle the challenge of efficient high-performance MIMO detection, we propose a non-convex homotopy optimization method for one-bit ML MIMO detection under binary symbol constellations.
	By extensive simulations, we show that the homotopy algorithm yields considerably better detection performance (specifically, bit-error rate performance) than some state-of-the-art algorithms.
	However it has a relatively high complexity requirement, compared to the state-of-the-art.
	
	\item We apply deep unfolding to the proposed homotopy algorithm to learn a better algorithm.
	Simulation results show that the deep-unfolded homotopy algorithm has detection performance comparable to the original homotopy algorithm, and it does so with a much lower complexity requirement---20 layers, or iterations, in all of our tested MIMO settings.
	The deep-unfolded homotopy algorithm also runs faster than the state-of-the-art algorithms.
	
	\item While our initial focus is on one-bit MIMO detection, the proposed homotopy method is also applicable to classical (unquantized) MIMO detection.
	Simulation results show that the homotopy algorithms (deep and non-deep) provide near-optimal detection performance, and
	the other empirical observations are similar to those in the one-bit case.

\end{enumerate}

We should discuss related works.
In the decades of MIMO detection research,
non-convex optimization was considered; see, e.g., \cite{blunt2005iterative,liu2017discrete}.
But the notion of homotopy optimization for attempting to avoid poor local minima, as well as our non-convex continuous reformulation for binary optimization, appear to have not been previously applied to MIMO detection.
Also, our homotopy method for binary optimization appears to have not been seen in the prior homotopy optimization literature.
Furthermore, the majority of the current deep MIMO detection studies consider the classical case.
Our deep unfolding design covers both the classical and one-bit MIMO cases.

For the sake of reproducible research, we have released the source code at\\
\texttt{\url{http://www.ee.cuhk.edu.hk/~wkma/mimo/}}
or at
\texttt{\url{https://github.com/mjshao-cuhk/DeepHOTML}}.


\section{Problem Statement}

\subsection{Model and ML Detection}

Let us begin by posing our problem in its generic abstract form.
Consider a data model
\beq \label{eq:model}
\by = \sgn( \bH \bx + \bv ),
\eeq
where $\by \in \{ -1, 1 \}^M$ is an observed data vector;
$\bx \in \{ -1, 1 \}^N$ is a binary vector;
$\sgn$ denotes the element-wise sign function;
$\bH  \in \Rbb^{M \times N}$ is a system matrix;
$\bv \in \Rbb^M$ is element-wise independent and identically distributed (i.i.d.) Gaussian noise, with mean zero and variance $\sigma^2$.
Our task is to detect $\bx$ from $\by$, given that $\bH$ and $\sigma$ are known.
We do so by pursuing the maximum-likelihood (ML) detection approach---which guarantees the minimum error probability of detecting $\bx$ under the assumption of element-wise  uniform i.i.d. $\bx$.
Under the model \eqref{eq:model}, the ML detector takes the form
\beq \label{eq:prob_main}
\hat{\bx}_{\rm ML} = \arg \min_{\bx \in \{ -1, 1 \}^N } f(\bx) \triangleq - \sum_{i=1}^M \log \Phi \left(  \frac{ y_i \bh_i^T \bx }{ \sigma } \right),
\eeq
where $\Phi(t) = \int_{-\infty}^t \frac{1}{\sqrt{2\pi}} e^{-\tau^2/2} d\tau$, and
$\bh_i$ denotes the $i$th row of $\bH$;
see~\cite{choi2015quantized,choi2016near}.
The ML problem \eqref{eq:prob_main} is a discrete optimization problem due to the binary constraint $\bx \in \{ -1 , 1\}^N$,
and a method capable of finding $\hat{\bx}_{\rm ML}$ tractably---given any instance $(\by,\bH,\sigma)$---appears to be unavailable.
In fact, it is computationally challenging to solve problem \eqref{eq:prob_main} exactly when the problem dimension $N$ is large;
e.g., complete enumeration of all points in $\{ -1, 1 \}^N$ is impossible for large $N$.
Our problem is to find an efficient strategy to tackle problem \eqref{eq:prob_main}.

The above problem arises in one-bit massive MIMO detection which has spurred great interest recently.
To describe it, consider an uplink multiuser MIMO scenario where multiple single-antenna users simultaneously transmit information symbols to a massive MIMO base station (BS);
and, to reduce the hardware cost, the BS
employs low-resolution ADCs.
Assuming flat fading channels, the signal model of the
above
scenario can be formulated as
\beq \label{eq:model_complex}
\by_{\rm C} = \mathcal{Q}( \bH_{\rm C} \bx_{\rm C} + \bv_{\rm C} ),
\eeq
where
$\by_{\rm C} \in \Cbb^{M_{\rm C}}$ is a receive vector whose $i$th element $y_{{\rm C},i}$ represents the complex baseband received signal at the $i$th antenna of the BS;
$\bx_{\rm C} \in \setX_{\rm C}^{N_{\rm C}}$ is a transmit vector whose $i$th element $x_{{\rm C},i}$ is the symbol transmitted by the $i$th user;
$M_{\rm C}$ and $N_{\rm C}$ are the numbers of receive antennas and users, respectively;
$\setX_{\rm C} \subset \Cbb$ is the symbol constellation set;
$\bH_{\rm C} \in \Cbb^{M_{\rm C} \times N_{\rm C}}$ is the MIMO channel;
$\mathcal{Q}$ denotes an element-wise quantization function;
$\bv_{\rm C} \in \Cbb^{M_{\rm C}}$ is element-wise i.i.d. circular Gaussian noise with mean zero and variance $\sigma_{\rm C}^2$.
Let us consider the case of one-bit quantization $\mathcal{Q}(\by_{\rm C}) = \sgn(\Rfrak\{ \by_{\rm C} \}) + \jj \cdot \sgn(\Ifrak\{ \by_{\rm C} \})$.
Also we focus on an in-phase quadrature-phase binary constellation $\setX_{\rm C} = \{ \pm 1 \pm \jj \}$, which is $4$-ary quadratic amplitude modulation (QAM) or quaternary phase shift keying (QPSK) constellation.
Under the aforementioned one-bit quantization and binary constellation,
the complex-valued model \eqref{eq:model_complex} can be rewritten as the real-valued model \eqref{eq:model} by setting $(M,N)= (2M_{\rm C},2N_{\rm C})$, $\sigma^2 = \sigma_{\rm C}^2/2$,
\ifconfver
	\beq \label{eq:c2r}
	\begin{gathered}
	\by = \begin{bmatrix}
	\Rfrak(\by_{\rm C}) \\ \Ifrak(\by_{\rm C})
	\end{bmatrix},
	\bx = \begin{bmatrix}
	\Rfrak(\bx_{\rm C}) \\ \Ifrak(\bx_{\rm C})
	\end{bmatrix},
	\bv = \begin{bmatrix}
	\Rfrak(\bv_{\rm C}) \\ \Ifrak(\bv_{\rm C})
	\end{bmatrix},
	\\
	\bH = \begin{bmatrix}
	\Rfrak(\bH_{\rm C}) & -\Ifrak(\bH_{\rm C})  \\ \Ifrak(\bH_{\rm C}) & \Rfrak(\bH_{\rm C})
	\end{bmatrix}.
	\end{gathered}
	\eeq
\else
	\beq \label{eq:c2r}
	\by = \begin{bmatrix}
	\Rfrak(\by_{\rm C}) \\ \Ifrak(\by_{\rm C})
	\end{bmatrix},
	\bx = \begin{bmatrix}
	\Rfrak(\bx_{\rm C}) \\ \Ifrak(\bx_{\rm C})
	\end{bmatrix},
	\bv = \begin{bmatrix}
	\Rfrak(\bv_{\rm C}) \\ \Ifrak(\bv_{\rm C})
	\end{bmatrix},
	\bH = \begin{bmatrix}
	\Rfrak(\bH_{\rm C}) & -\Ifrak(\bH_{\rm C})  \\ \Ifrak(\bH_{\rm C}) & \Rfrak(\bH_{\rm C})
	\end{bmatrix}.
	\eeq
\fi
By the above relation,
the ML detector \eqref{eq:prob_main} applies to the one-bit MIMO model \eqref{eq:model_complex}.

\subsection{Some Aspects}

Some basic nature of the ML problem \eqref{eq:prob_main} should be noted.
First,
the objective function $f$ of problem \eqref{eq:prob_main} is convex (this is mainly because $-\log \Phi(x)$ is convex~\cite{boyd2004convex}).
The convexity of $f$ can be leveraged to develop efficient approximate ML detectors.
In \cite{choi2016near}, the authors considered a convex sphere relaxation of the ML problem \eqref{eq:prob_main};
specifically,
\beq \label{eq:sr}
\min_{\| \bx \|^2 \leq N} f(\bx),
\eeq
where $\| \cdot \|$ denotes the Euclidean norm.
Second, it is natural to question whether we can reliably perform MIMO detection from one-bit observations.
From the one-bit model \eqref{eq:model} and its ML detector \eqref{eq:prob_main},
it is intuitively unobvious why or whether the true vector $\bx$ can be recovered from the heavily quantized observation $\by$.
We provide a clue to this basic question by the following simple result.
\begin{Fact} \label{fact:1}
	Suppose that
	\begin{enumerate}[(a)]
	\item $\bh_1,\ldots,\bh_M$ are i.i.d.;
	\item the probability density function of the $\bh_i$'s, denoted by $q(\bh)$, is continuous on its support;
	\item  the support of $q(\bh)$ is $\Rbb^N$.
	\end{enumerate}	
	Consider $M \rightarrow \infty$ such that
	\beq \label{eq:tf}	
	\frac{1}{M} f (\bx) \xrightarrow{P} \tilde{f}(\bx) \triangleq - \Exp\left[ \log \Phi\left( \frac{y \bh^T \bx }{{\sigma}} \right)   \right],
	\eeq
	provided the expectation exists. Here,
	 $\xrightarrow{P}$ denotes convergence in probability;
	$y$ and $\bh$ represent random variables associated with the realizations $y_i$'s and $\bh_i$'s, respectively;
	the expectation $\Exp[ \cdot ]$ is taken with respect to $y$ and $\bh$.
	Then,
	the minimizer of $\tilde{f}(\bx)$ over $\Rbb^N$ is uniquely given by the true binary vector in the signal model~\eqref{eq:model}.
	It follows that the ML problem
	\[
	\min_{\bx \in \{ -1, 1 \}^N } \tilde{f}(\bx)
	\]
	has its solution uniquely given by the true binary vector.
\end{Fact}

The result in Fact~\ref{fact:1}
was first reported in  \cite[Lemma 2]{choi2015quantized} for the case of i.i.d. Gaussian $\bh$ and constant Euclidean-norm $\bx$, and here we show a more general result.
Before giving the proof, let us first discuss the implications.
Fact~\ref{fact:1} suggests that if the number of antennas at the BS is very large (true for massive MIMO), then the ML problem \eqref{eq:prob_main} may lead to correct recovery of the true binary vector.
In fact, this large-$M$ recovery result holds not only for the ML problem, but also for unconstrained relaxation of problem~\eqref{eq:prob_main} (i.e., replacing the constraint $\{ -1, 1 \}^N$ by $\Rbb^N$) and
 the sphere relaxation in \eqref{eq:sr};
it also holds if
 the true vector $\bx$ in the signal model \eqref{eq:model} is drawn from
a higher-order constellation set (or even $\Rbb^N$).
In addition, the requirement with the channel distribution is quite general; e.g., it works for i.i.d. Gaussian channels, correlated Gaussian channels (with respect to users), or other continuously distributed channels (with support $\Rbb^N$).

{\em Proof of Fact~\ref{fact:1}:} \
 Our proof is different from \cite[Lemma 2]{choi2015quantized}, which uses stochastic orders.
We employ a more basic proof, taking proof ideas from the consistency property of ML estimation.
To avoid notation overlap, re-denote the true binary vector in the signal model \eqref{eq:model} as $\bx_0$.
The probability mass function of $y_i$ given $\bh_i$ and $\bx_0$ is
\beq \label{eq:p_cond}
 p(y_i | \bh_i, \bx_0 ) = \Phi \left(  \frac{ y_i \bh_i^T \bx_0 }{ \sigma } \right),
\eeq
which can be shown from \eqref{eq:model}.
We see that, for any $\bx \neq \bx_0$,
\begin{align*}
& \tilde{f}(\bx_0) - \tilde{f}(\bx)  = \Exp\left[ \log\left( \frac{p(y| \bh,\bx)}{p(y| \bh,\bx_0)} \right) \right] \\
& =
\sum_{y \in \{-1,1\}} \int_{\Rbb^N}
\log\left( \frac{p(y| \bh,\bx)}{p(y| \bh,\bx_0)} \right) p(y| \bh,\bx_0) q(\bh) d\bh \\
& \leq \sum_{y \in \{-1,1\}} \int_{\Rbb^N}
\left( \frac{p(y| \bh,\bx)}{p(y| \bh,\bx_0)} - 1 \right) p(y| \bh,\bx_0) q(\bh) d\bh \\
& = 0,
\end{align*}
where the inequality is due to $\log(t) \leq t -1$.
The above inequality implies that $\bx_0$ is a minimizer of $\tilde{f}(\bx)$ over $\Rbb^N$.
To show that $\bx_0$ is the unique minimizer, note that equality in the above equation holds if and only if
$p(y| \bh,\bx) = p(y| \bh,\bx_0)$ for every $y \in \{-1,1\}$ and $\bh \in \Rbb^N$.
The latter condition implies
\[
\bh^T \bx = \bh^T \bx_0, \quad  \text{for every $\bh \in \Rbb^N$};
\]
this is seen from \eqref{eq:p_cond} (the monotonicity of $\Phi$ should also be noted).
Clearly the
above equation
does not hold for any $\bx \neq \bx_0$,
which means that there does not exist $\bx \neq \bx_0$ such that $\tilde{f}(\bx) = \tilde{f}(\bx_0)$.
The proof is complete.
\hfill $\blacksquare$

\section{A Homotopy Optimization Method}
\label{sect:homotopy}

\subsection{The Main Idea}

Our strategy for tackling the ML problem \eqref{eq:prob_main} hinges on a very recently introduced penalty approach~\cite{shao2019framework}.
Consider a penalty, and possibly approximate, formulation of problem \eqref{eq:prob_main}
\beq \label{eq:prob_pen}
({\rm P}_\lambda) \qquad  	\min_{\bx \in [-1, 1]^N } F_\lambda(\bx) \triangleq f(\bx) - \lambda \| \bx \|^2,
\eeq
for a given parameter $\lambda \geq 0$.
Intuitively, the idea is to encourage large value with every $x_i^2$ by imposing the penalty term $- \lambda \| \bx \|^2$ in the objective, on the one hand, and limit  $x_i^2 \leq 1$ on the other hand.
In doing so, the optimal solution to problem \eqref{eq:prob_pen} should be forced to $x_i^2 = 1$, or $x_i \in \{ -1 , 1 \}$, if we apply a large $\lambda$.
In fact, under a mild assumption, this intuition is correct:
\begin{Theorem}[a rephrased version of Theorem 2 in~\cite{shao2019framework}]
	Let $f$ be a twice differentiable function, not necessarily the one in \eqref{eq:prob_main},
	and consider the corresponding problems in \eqref{eq:prob_main} and \eqref{eq:prob_pen}.
	Let $L_f > 0$ be a Lipschitz constant of the gradient of $f$ on $[-1,1]^N$,
 which must exist for twice differentiable $f$.
	For any $\lambda > L_f/2$, it holds that\footnote{As a technical remark, Theorem 2 in \cite{shao2019framework} only showed statement (a) of Theorem 1. Statements (b)--(c) are straightforward corollaries of statement (a).}
	\begin{enumerate}[(a)]
		\item any locally optimal solution to problem \eqref{eq:prob_pen} lies in $\{ -1, 1 \}^N$;
		\item any globally optimal solution to problem \eqref{eq:prob_pen} is also that to problem \eqref{eq:prob_main};
		\item if $\bx$ is a stationary point of problem \eqref{eq:prob_pen} and $\bx$ does not lie in $\{-1, 1 \}^N$, then $\bx$ must be either a local maximum or a saddle point.		
	\end{enumerate}
\end{Theorem}
Hence, for a sufficiently large $\lambda$, we can employ problem \eqref{eq:prob_pen} as an equivalent formulation of the ML problem \eqref{eq:prob_main}.

As the main benefit,
the penalty formulation \eqref{eq:prob_pen}
turns the discrete ML problem into a continuous, and convex constrained, optimization problem.
As a result, we can use methods, such as the descent-based methods, to efficiently compute a stationary point of problem \eqref{eq:prob_pen} (and hopefully a locally optimal solution to it).
But we should also note that the penalty term $-\lambda \| \bx \|^2$ in problem \eqref{eq:prob_pen} makes the problem non-convex,
and a descent-based algorithm can converge to a poor local minimum.

\begin{algorithm}[htb!]
	\caption{A homotopy strategy for tackling problem \eqref{eq:prob_main}} \label{alg:homotopy}
	\begin{algorithmic}[1]		
		\STATE {\bf given} an initial penalty parameter $\lambda_0 \geq 0$ and a starting point $\bx^0$
		\STATE $k= 0$
		\REPEAT
		\STATE $k= k+1$
		\STATE  set $\lambda_k$ as increased version of $\lambda_{k-1}$
		\STATE  run a descent-based algorithm, with $\bx^{k-1}$ as the starting point, to compute a stationary point of problem $({\rm P}_{\lambda_k})$ in \eqref{eq:prob_pen}, and store the solution obtained as $\bx^k$
		\UNTIL{$\lambda_k$ is larger than a pre-specified threshold.   }
		\STATE {\bf output} $\bx^k$		
	\end{algorithmic}
\end{algorithm}

\begin{figure}[h]
	\centering
	\begin{subfigure}[b]{0.45\textwidth}
		\includegraphics[width=\textwidth]{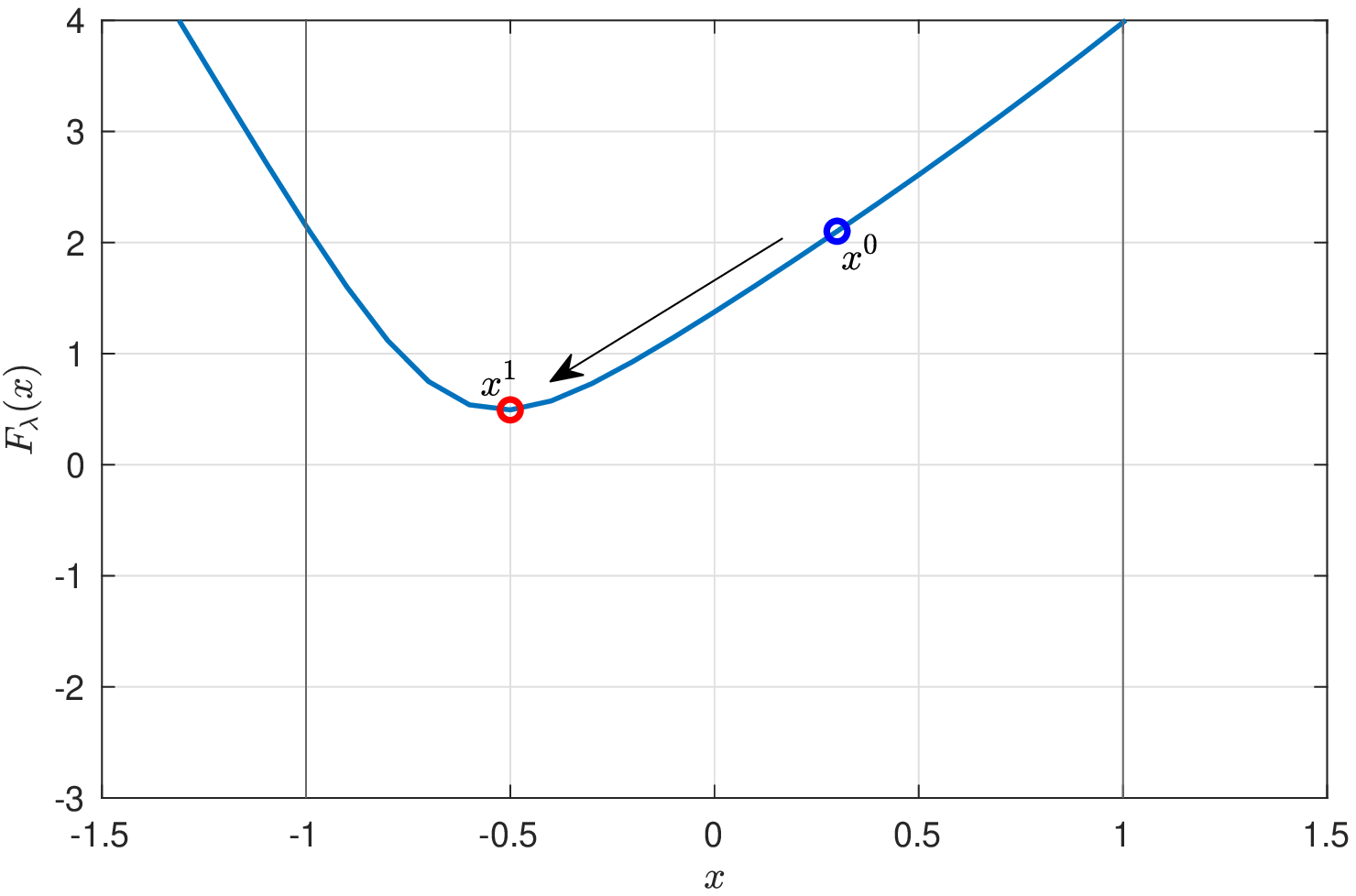}
		\caption{from ${\bx}^0$ to $\bx^1$, $\lambda= \lambda_1$}
	\end{subfigure}
	~ 
	\begin{subfigure}[b]{0.45\textwidth}
		\includegraphics[width=\textwidth]{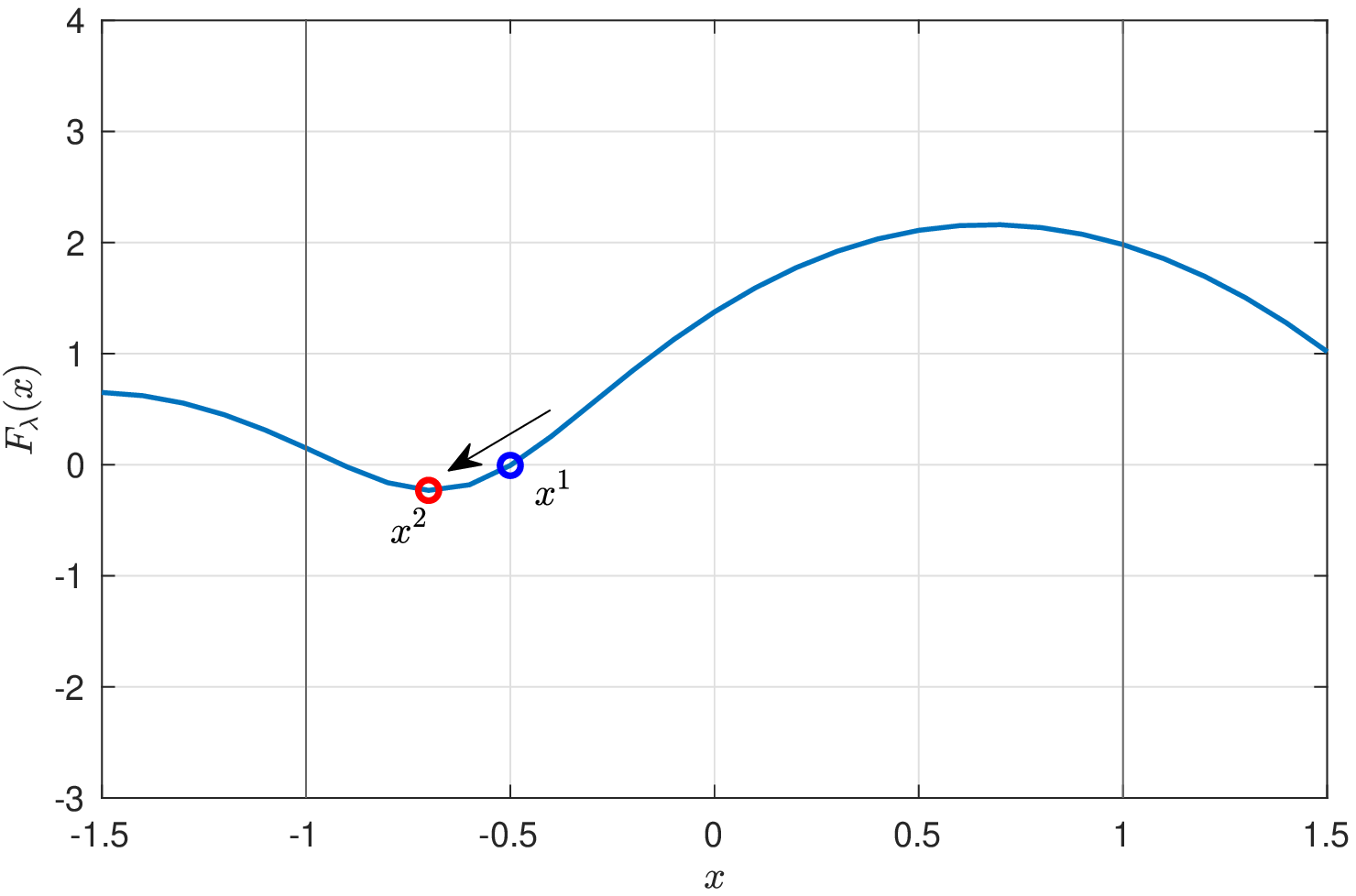}
		\caption{from ${\bx}^1$ to $\bx^2$, $\lambda= \lambda_2$}
	\end{subfigure}
	~ 
	\begin{subfigure}[b]{0.45\textwidth}
		\includegraphics[width=\textwidth]{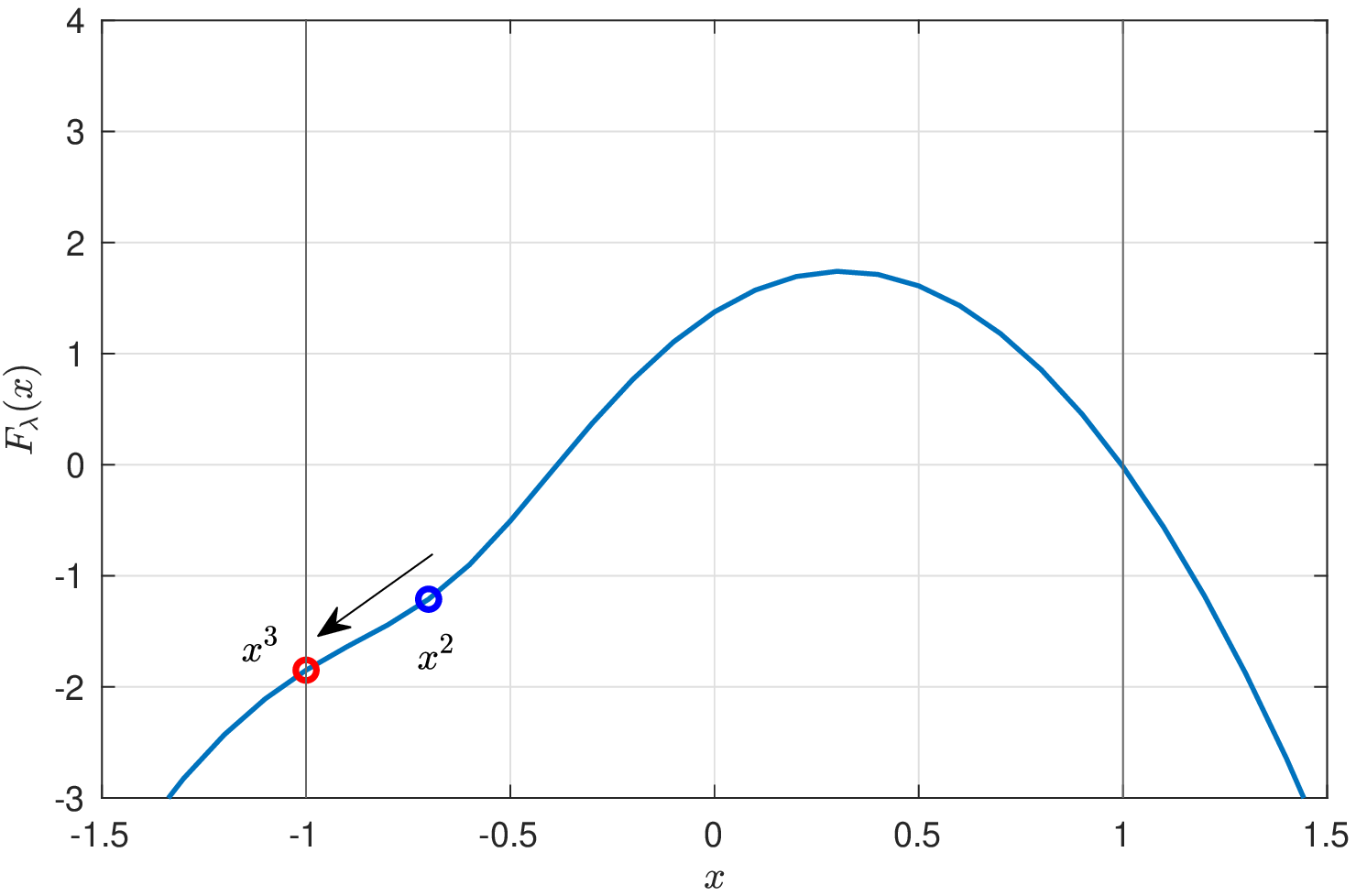}
		\caption{from ${\bx}^3$ to $\bx^2$, $\lambda= \lambda_3$}
	\end{subfigure}
	
	\caption{Illustration of how the homotopy method works.}\label{fig:lambda_1D}
\end{figure}

As an attempt to circumvent local minima, we consider the following strategy.
Problem \eqref{eq:prob_pen} for $\lambda = 0$, or problem $({\rm P}_0)$, is convex.
This gives an intuition that problem $({\rm P}_\lambda)$ in \eqref{eq:prob_pen} should be easy for small $\lambda$, and hard for large $\lambda$.
It is therefore natural to consider the optimization strategy in Algorithm~\ref{alg:homotopy},
where we progressively increase $\lambda$, or the difficulty of the problem.
Also, and just as important, we use the previous solution $\bx^{k-1}$ in an effort to find a good solution to problem $({\rm P}_{\lambda^k})$.
Imagine this:
If $\bx^{k-1}$ is a globally optimal solution to problem $({\rm P}_{\lambda_{k-1}})$ and the change of $\lambda_k$ relative to $\lambda_{k-1}$ is small,
then the optimization landscape may undergo only mild changes, and a descent-based algorithm starting with $\bx^{k-1}$ may ``easily'' descend to a close-by globally optimal solution to problem $({\rm P}_{\lambda_k})$.
Thus we may be able to find the ML solution by tracing a solution path of problem $({\rm P}_{\lambda})$, from small $\lambda$ to large $\lambda$.
We give the reader more insight by pictorially illustrating the above described idea in Fig.~\ref{fig:lambda_1D}, and by displaying the landscape of a 2D instance of problem $({\rm P}_{\lambda})$ in Fig.~\ref{fig:lambda_2D}.

\ifconfver
	\begin{figure*}[thb]
\else
	\begin{figure}[thb]
\fi
	\centering
	\begin{subfigure}[b]{\customfigwidth}
		\includegraphics[width=\textwidth]{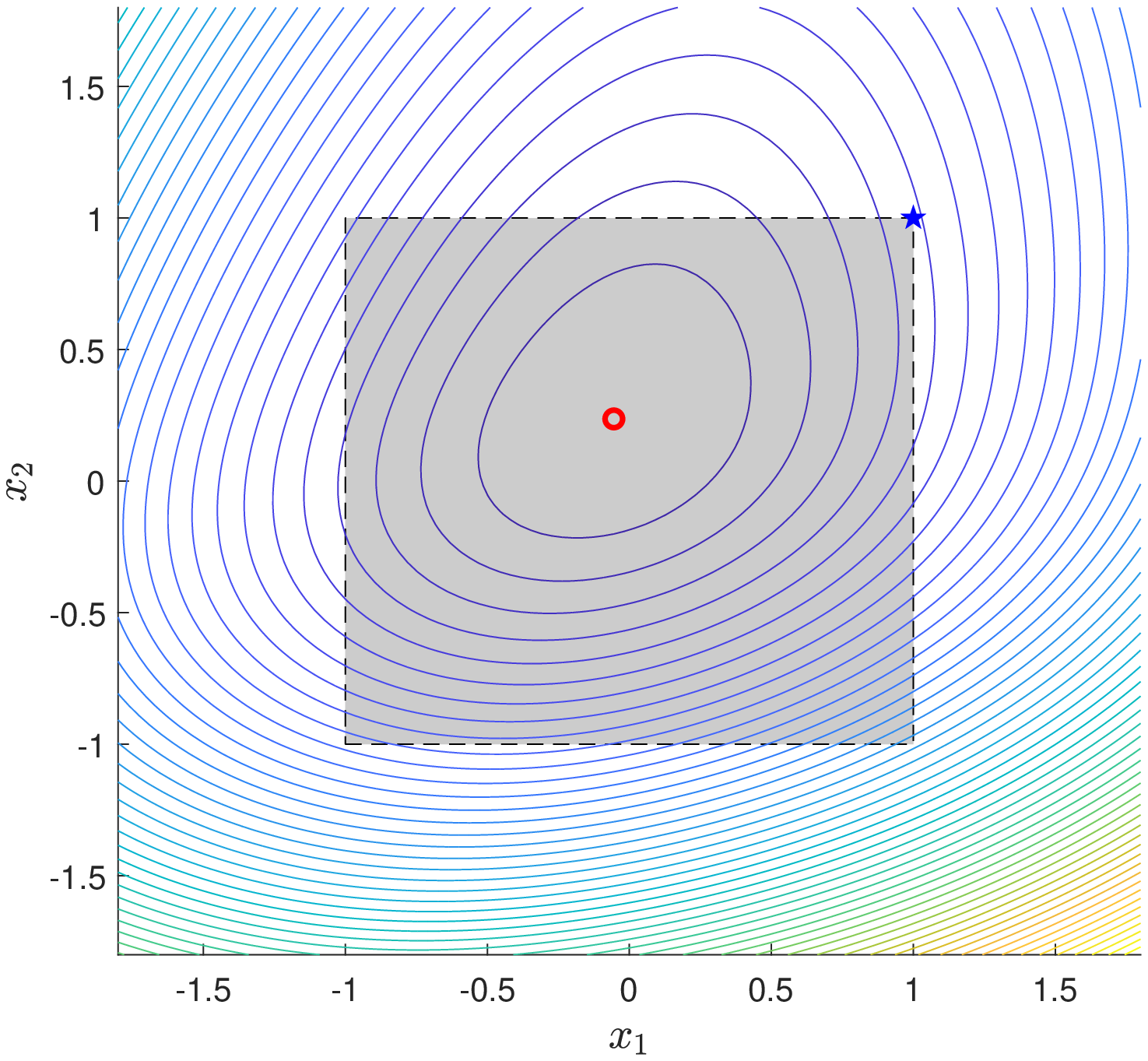}
		\caption{$\lambda=0$}\label{subfig:lambda00}
	\end{subfigure}
	~ 
	\begin{subfigure}[b]{\customfigwidth}
		\includegraphics[width=\textwidth]{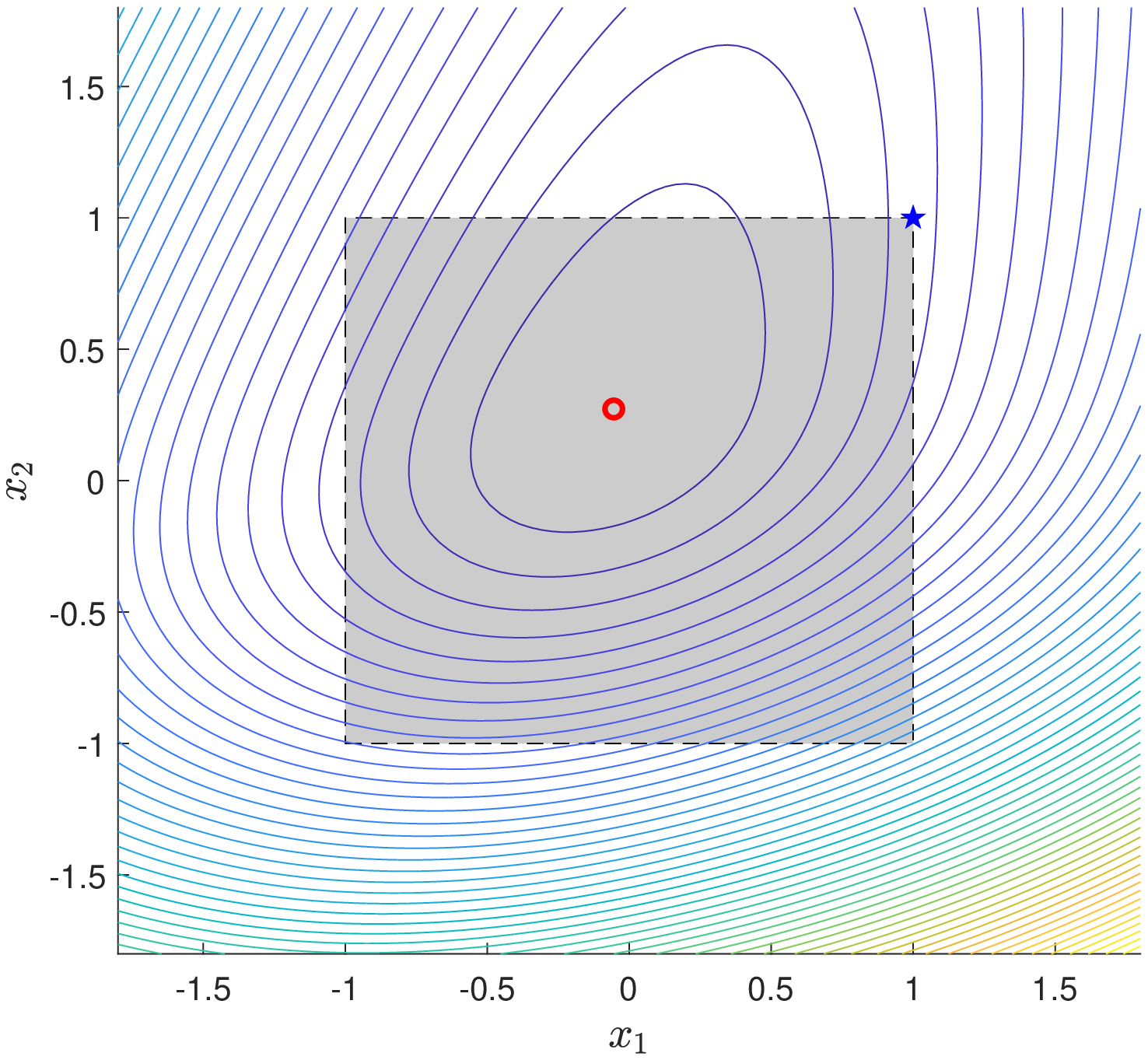}
		\caption{$\lambda=2$}
	\end{subfigure}
	~ 
	\begin{subfigure}[b]{\customfigwidth}
		\includegraphics[width=\textwidth]{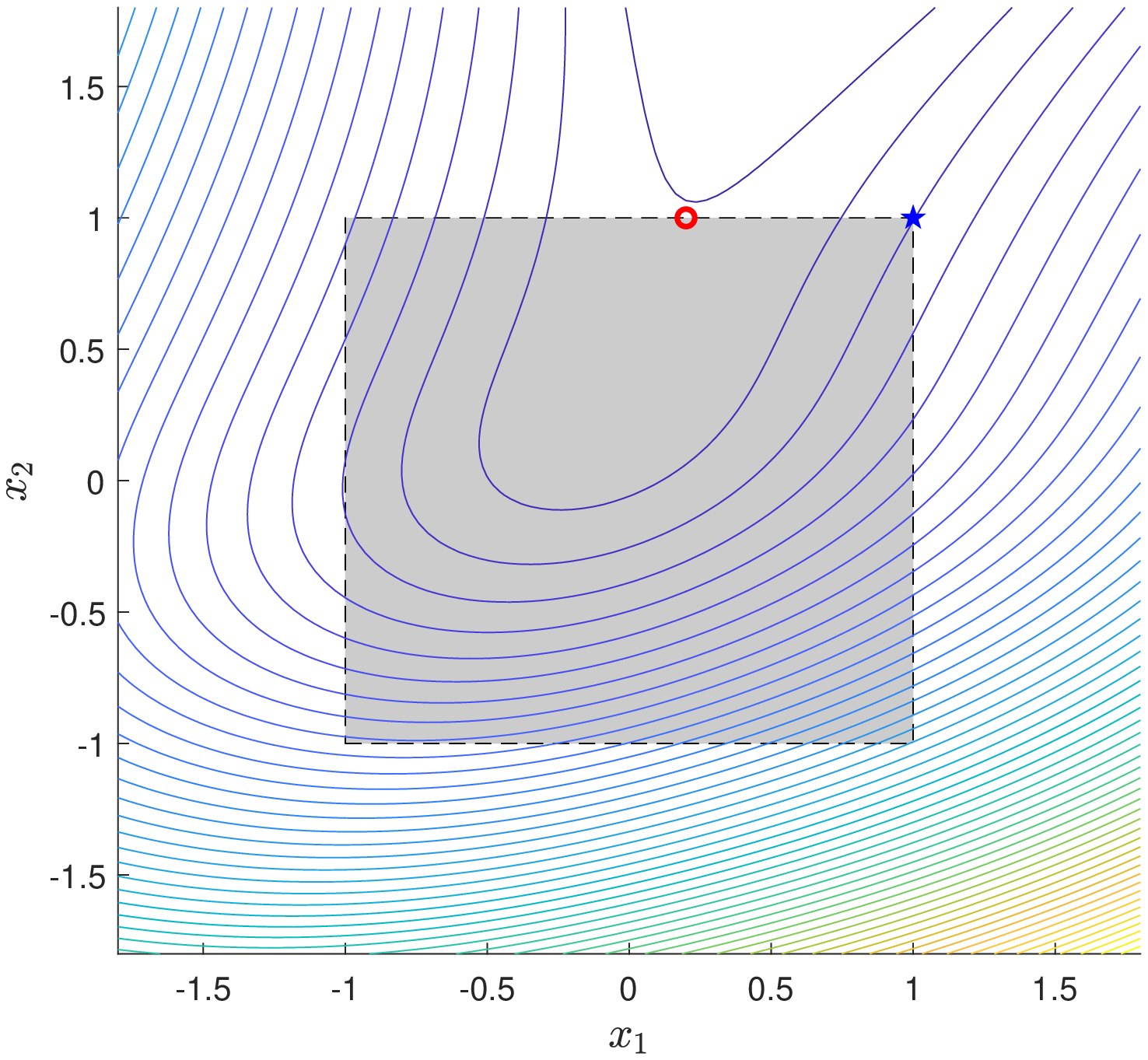}
		\caption{$\lambda=4$}
	\end{subfigure}
	~ 
	\begin{subfigure}[b]{\customfigwidth}
		\includegraphics[width=\textwidth]{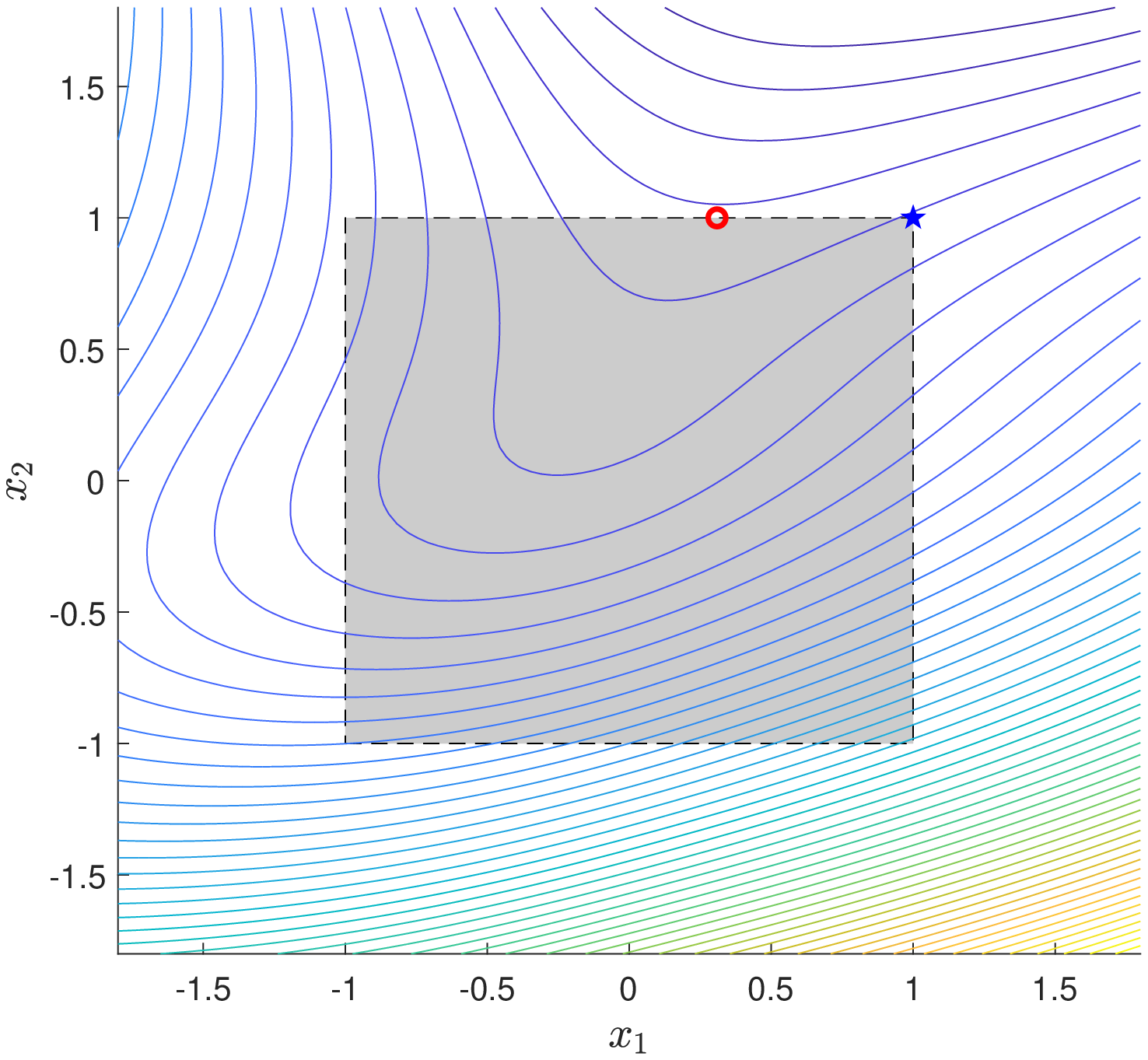}
		\caption{$\lambda=6$}
	\end{subfigure}
	~ 
	\begin{subfigure}[b]{\customfigwidth}
		\includegraphics[width=\textwidth]{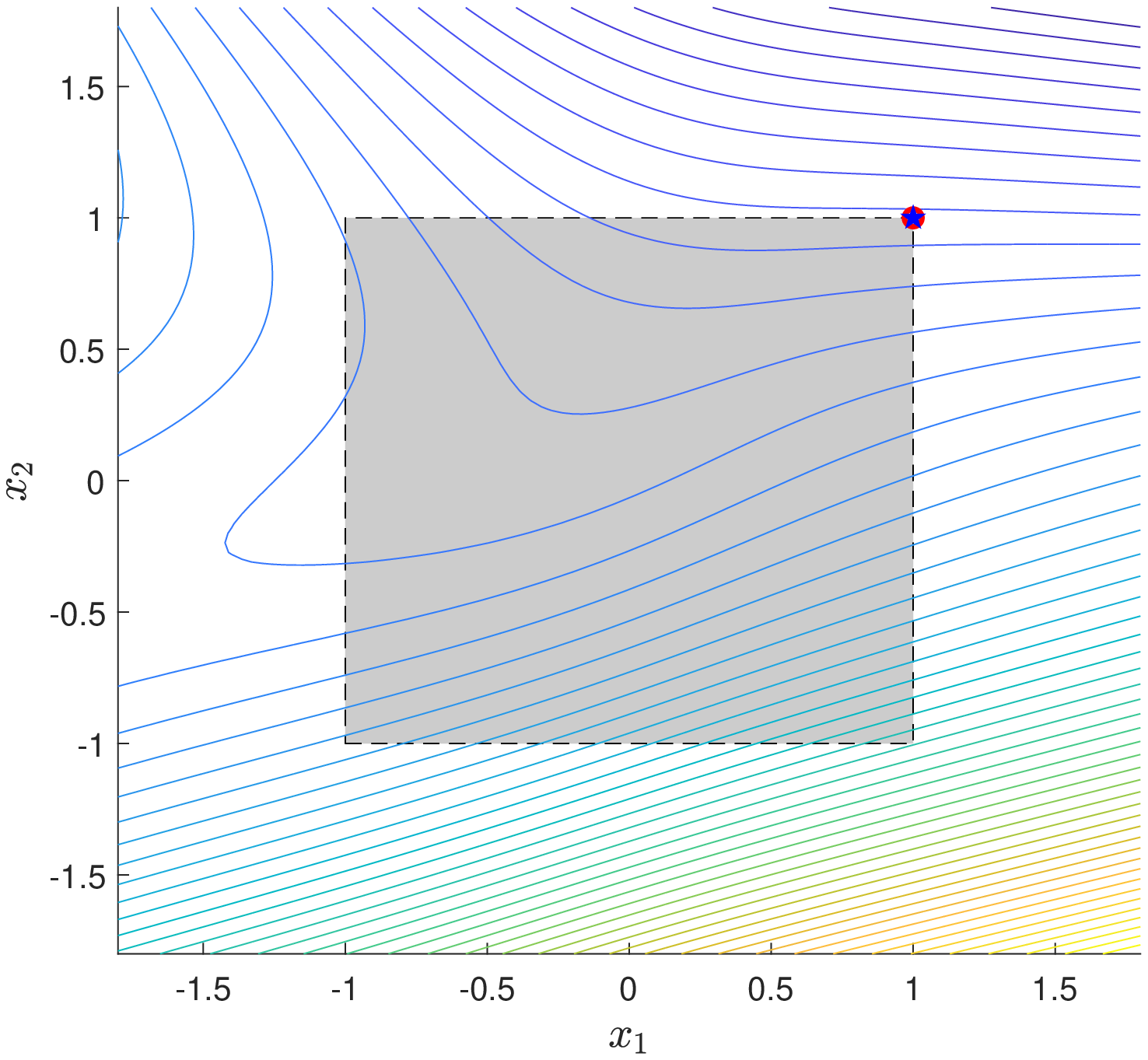}
		\caption{$\lambda=8$}\label{subfig:lambda8}
	\end{subfigure}
	\begin{subfigure}[b]{\customfigwidth}
		\includegraphics[width=\textwidth]{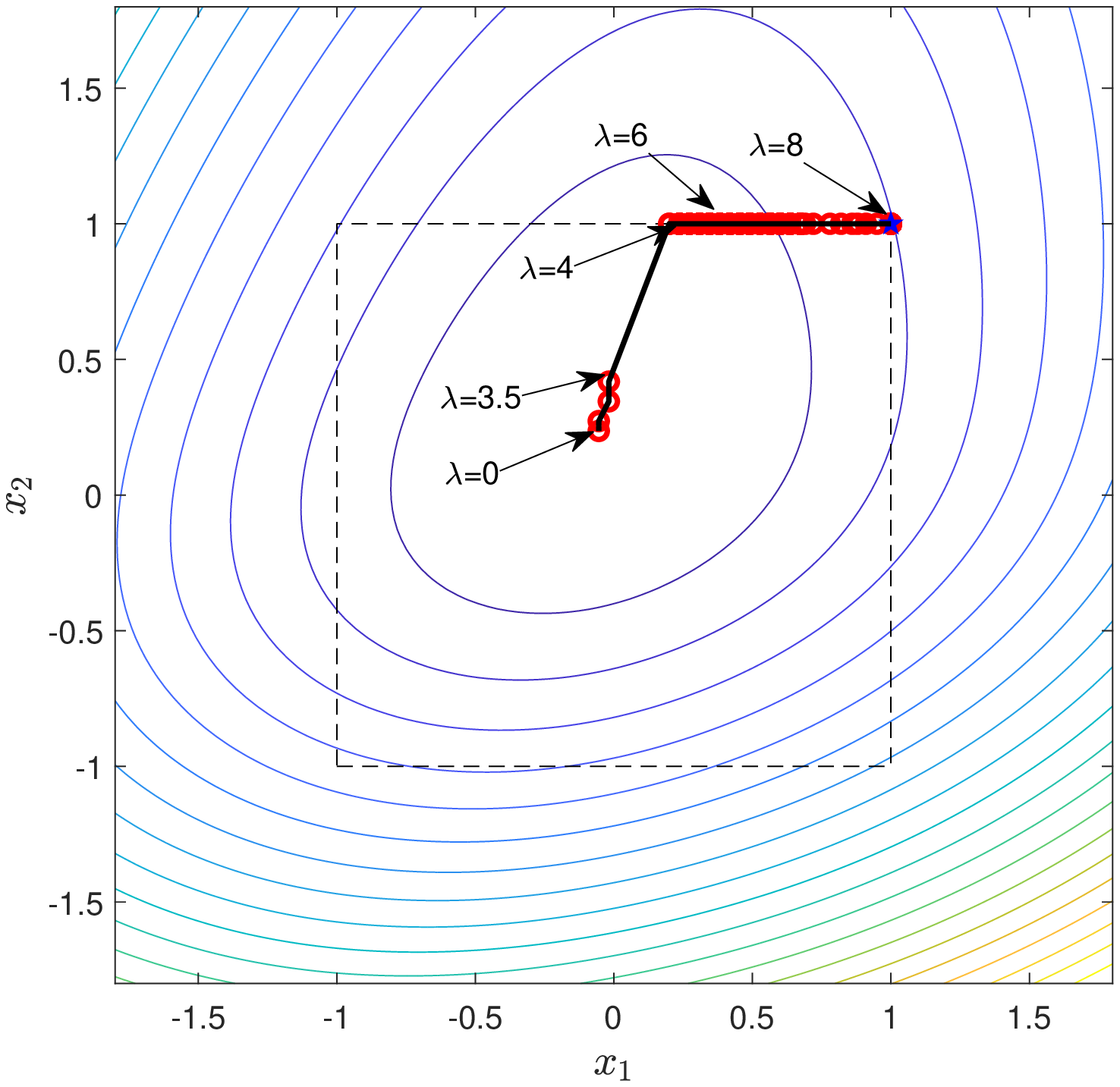}
		\caption{solution path}\label{fig:lambdapath}
	\end{subfigure}
	\caption{An instance of the landscape of problem $({\rm P}_{\lambda})$ in \eqref{eq:prob_pen}, with $N=2$. The contour is $F_{\lambda}(\bx)$; the gray area is the feasible set; the red circle is the optimal solution to problem $({\rm P}_{\lambda})$; the blue star is the optimal ML solution;
	the instance is a randomly generated one for one-bit MIMO detection, with $M= 10, \sigma=1$.}\label{fig:lambda_2D}
\ifconfver
	\end{figure*}
	\else
	\end{figure}
\fi

\begin{figure}[ht!]
	\centering
	\includegraphics[width=.45\textwidth]{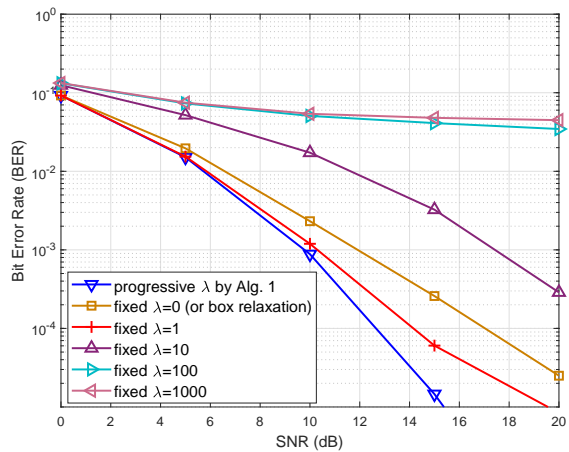}
	\caption{ Performance of using formulation \eqref{eq:prob_pen} to tackle the one-bit ML MIMO detection problem.}\label{fig:demo_ber}
\end{figure}

To demonstrate the benefits of the above optimization strategy, we show a simulation result here.
We consider one-bit MIMO detection with $(M,N)= (256,48)$.
We perform detection by tackling the ML problem \eqref{eq:prob_main} via  formulation  \eqref{eq:prob_pen}, either with a fixed $\lambda$ or with the progressively increasing $\lambda$ strategy in Algorithm~\ref{alg:homotopy}.
The results are shown in Fig.~\ref{fig:demo_ber}.
We observe that the progressively increasing $\lambda$ strategy leads to better bit-error rate performance than fixing $\lambda$.
In particular, using a large fixed $\lambda$  yields unsatisfactory results, most likely due to convergence to poor local minima.
Using a smaller fixed $\lambda$ mitigates the undesirable effects, but doing so also weakens its approximation accuracy relative to the ML.

Algorithm 1 may be taxonomized into the class of homotopy optimizaton methods;
see
\cite{wu1996effective,dunlavy2005homotopy,xiao2013proximal,mobahi2015link,hazan2016graduated,anandkumar2017homotopy} and the references therein.
The principle of homotopy optimization is to first find a transformation, or a homotopy map, that maps the original problem to another problem.
That transformation depends on a parameter, say, $\lambda$.
For large $\lambda$, the transformed problem is close to the original problem but is hard to solve directly.
For small $\lambda$, the transformed problem is easy to solve but poorly approximates the original problem.
Then, we seek to find a solution (possibly approximate) to the original problem by attempting to trace the solution path of a sequence of such transformed problems, from easy to hard and in a gradually changing fashion, e.g., by the routine we saw in  Algorithm 1 which is typical in homotopy methods.

\begin{Remark}[Related Work]
	We close this subsection by discussing the relationship of the proposed homotopy formulation and the existing methods.
	For $\lambda= 0$, formulation \eqref{eq:prob_pen} reduces to convex box relaxation  \cite{tan2001constrained,yener2002cdma,thrampoulidis2018symbol}.
	Thus, formulation \eqref{eq:prob_pen} may be regarded as a non-convex enhancement of box relaxation.
	Moreover, the negative square penalty $-\lambda \| \bx \|^2$ used in formulation \eqref{eq:prob_pen} also appears in other contexts, such as low-density parity-check decoding \cite{Liu2016ADMM} and one-bit precoding \cite{Castaneda2017massive}, to force the solution closer to a binary vector.
	A subtle but important difference is that
	the aforementioned studies often employ a fixed penalty parameter $\lambda$,
	while we use the homotopy optimization strategy to progressively adjust $\lambda$.
\end{Remark}

\subsection{How to Choose the Penalty Sequence $\{ \lambda_k \}$?}
\label{sect:homotopy_ldr}

In the homotopy strategy in Algorithm 1,
a crucial question is how the penalty parameter sequence $\{ \lambda_k \}$ should be chosen.
Having a small increase with $\lambda_k$ at each iteration may allow us to follow the solution path better, but it may take too many iterations to do so.
Having a large increase with $\lambda_k$, on the other hand,
reduces the number of iterations
but
may also increase the risk of losing track of the solution path.
By our experience, a heuristically chosen $\{ \lambda_k \}$ works well---empirically.
For example, we may choose $\lambda_k = \lambda_{k-1} c$ for some constant $c > 1$.
But it is tempting to see whether some theory-guided selection rule exists.
Here we show one such rule by drawing a connection with Lagrangian dual relaxation (LDR).

We start with studying an equivalent formulation of the ML problem \eqref{eq:prob_main}
\beq \label{eq:prob_main_LDR}
\begin{aligned}
f^\star = \min_{\bx \in \setD} & ~ f(\bx) \\
{\rm s.t.} & ~ \| \bx \|^2 \geq N
\end{aligned}
\eeq
where $\setD= [ -1, 1 ]^N$ denotes the problem domain, and $f^\star$ the optimal value.
We are interested in the Lagrangian dual of the formulation in \eqref{eq:prob_main_LDR}.
Let
\[
L(\bx,\lambda) = f(\bx) + \lambda( N - \| \bx \|^2 )
\]
be the Lagrangian of problem \eqref{eq:prob_main_LDR}, and let
\beq \label{eq:d_lam}
d(\lambda) = \min_{\bx \in \setD} L(\bx,\lambda)
\eeq
be the dual function.
The dual problem of problem \eqref{eq:prob_main_LDR} is
\beq \label{eq:prob_dual_LDR}
d^\star = \max_{\lambda \geq 0} ~ d(\lambda)
\eeq
where $d^\star$ denotes the optimal value of problem \eqref{eq:prob_dual_LDR}.
At this point, note that the minimization problem in  \eqref{eq:d_lam} is equivalent to the penalty formulation \eqref{eq:prob_pen}.
Also, recall from the basic concepts of Lagrangian duality that the dual problem \eqref{eq:prob_dual_LDR} is convex, and the weak duality inequality $f^\star \geq d^\star$  holds.

LDR is a method that exploits the dual problem to approximate the original problem; see, e.g., \cite{poljak1995recipe,pan2013mimo}.
It works by first finding the optimal solution to the dual problem \eqref{eq:prob_dual_LDR}, denoted herein by $\lambda^\star$, and then finding the solution $\bx$ to the minimization problem in \eqref{eq:d_lam} for $\lambda = \lambda^\star$.
The duality gap $f^\star - d^\star$ provides strong indication of how well LDR approximates the original problem; the smaller the gap is, the better LDR should be.
For the problem at hand, we show that the LDR in \eqref{eq:prob_dual_LDR} is tight.
\begin{Theorem}
	Consider the same problem settings in Theorem 1. The primal-dual problem pair  \eqref{eq:prob_main_LDR} and \eqref{eq:prob_dual_LDR} achieves zero duality gap $f^\star - d^\star = 0$.
\end{Theorem}

{\em Proof:} \
Let $\bar{\lambda}$ be any constant such that $\bar{\lambda} > L_f /2$. We have $d^\star \geq d(\bar{\lambda})= f^\star$, where the equality above is due to Theorem 1.
This, together with $f^\star \geq d^\star$, implies that $f^\star = d^\star$. \hfill $\blacksquare$
\medskip

There is a connection between the preceding homotopy strategy and the LDR here.
Consider finding the optimal solution to the dual problem \eqref{eq:prob_dual_LDR} via the projected subgradient method \cite{boyd2003subgradient}, given by
\beq \label{eq:subgrad_iter}
\lambda_k = \max\{ 0, \lambda_{k-1} + \mu_k g(\lambda_{k-1}) \}, \quad k=1,2,\cdots
\eeq
Here,
$g(\lambda)$ denotes a subgradient of the dual function $d$ at $\lambda$;
$\mu_k$ is a step size (see the literature \cite{boyd2003subgradient} for step-size rules that guarantee that $\lambda_k$ will converge to an optimal solution);
$\lambda_0$, the starting point, is assumed to be non-negative.
By the subgradient calculus \cite{boyd2003subgradient}, $g(\lambda)$ is given by
\beq \label{eq:subgrad_g}
g(\lambda) = N - \| \hat{\bx}_\lambda \|^2,
\eeq
where
\beq \label{eq:subgrad_hatx}
\hat{\bx}_\lambda \in \arg \min_{\bx \in \setD} L(\bx,\lambda).
\eeq
Putting \eqref{eq:subgrad_g}--\eqref{eq:subgrad_hatx} into \eqref{eq:subgrad_iter},
we can simplify the  projected subgradient method \eqref{eq:subgrad_iter} to
\beq \label{eq:subgrad_iter2}
\lambda_k =  \lambda_{k-1} + \mu_k (N- \| \hat{\bx}_{\lambda_{k-1}} \|^2), \quad k=1,2,\cdots
\eeq
where we have used $\lambda_0 \geq 0$ and $\| \bx \|^2 \leq N$ for any $\bx \in \setD$.
The above projected subgradient method is seen to be closely related to the homotopy strategy in Algorithm 1:
the former also increases $\lambda_k$ at each iteration, and it requires us to solve problem \eqref{eq:subgrad_hatx}---an equivalent of the penalty formulation \eqref{eq:prob_pen}---at each iteration.
The challenge with realizing the projected subprojected method is that problem \eqref{eq:subgrad_hatx} is non-convex.
But we can substitute the globally optimal solution $\hat{\bx}_{\lambda_k}$ with an approximate solution, obtained by using $\hat{\bx}_{\lambda_{k-1}}$ to warm-start a descent-based algorithm.
The resulting (approximate) projected subgradient method will then be identical to the homotopy strategy in Algorithm 1, with a clearly defined penalty update rule as shown in \eqref{eq:subgrad_iter2}.

Let us summarize. In the homotopy strategy in Algorithm 1, we may choose $\lambda_k$ by
\beq \label{eq:update_lambda}
\lambda_k =  \lambda_{k-1} + \mu_k (N- \| {\bx}^{k-1} \|^2),
\eeq
where $\{ \mu_k \}$ is a standard step-size sequence in the subgradient method.
It is interesting to observe that if $\bx^{k-1}$ is closer to $\{ -1, 1 \}^N$, which implies a smaller $N- \| {\bx}^{k-1} \|^2$,
then the increase of $\lambda_k$ will be slowed down---which seems to make sense intuitively.

\subsection{The Descent-Based Algorithm Used}
\label{sect:gemm}

We complete our design by specifying the descent-based algorithm in line 6 of Algorithm 1.
The constraint of problem \eqref{eq:prob_pen} has a simple structure,
and one can exploit such structure for efficient optimization via structured optimization methods---such as the projected gradient method.
Here we employ a more advanced version of the projected gradient method which was numerically found to be very fast in a different application~\cite{shao2019framework}.

The aforementioned method, called {\em gradient extrapolated majorization-minimization} (GEMM),
entails concepts from majorization-minimization and the accelerated project gradient method.
Here we concisely state the method, and the reader is referred to~\cite{shao2019framework} for detailed descriptions such as its intuitions and stationary-point convergence guarantee.
Let $\nabla f(\bx)$ denote the gradient of a differentiable function $f$ at $\bx$.
Given a starting point $\bx^0$,
the GEMM method iteratively runs
\beq \label{eq:gemm}
\bx^{t+1}= \Pi\left(  \bz^t + \beta_t \nabla G_\lambda(\bz^t | \bx^t )  \right), \quad t=0, 1,2,\cdots
\eeq
Here, $\Pi$ denotes the projection onto $[-1,1]^N$, which has a closed form $\Pi(\bx) = [~ \Pi(x_1), \ldots, \Pi(x_N) ~]^T$,
\[
\Pi(x) = \left\{
\begin{array}{ll}
-1, & x < -1 \\
x, & -1 \leq x \leq 1 \\
1, & x > 1
\end{array} \right.;
\]
$G_\lambda(\cdot|\bar{\bx})$ is a convex differentiable majorant of $F_\lambda$ at $\bar{\bx}$;
$\nabla G_\lambda(\bx|\bar{\bx})$ is the gradient of $G_\lambda(\cdot|\bar{\bx})$ at $\bx$;
$\beta_t$ is a step size;
$\bz^t$ is an extrapolated point of $\bx^t$ and takes the form
\begin{equation}\label{eq:FISTA}
\bz^t = \bx^t + \alpha_t (\bx^t - \bx^{t-1} )
\end{equation}
for some pre-specified extrapolation sequence $\{ \alpha_t \}$ (note that $\bx^{-1} = \bx^0$).
Simply speaking, the GEMM method in \eqref{eq:gemm} is an inexact majorization-minimization scheme that uses a one-iteration accelerated projected gradient update as its inexact update.
It reduces to the projected gradient method if we set $\alpha_t = 0$ and replace $\nabla G_\lambda(\cdot| \bx^t)$ with $\nabla F_\lambda(\cdot)$, i.e., no extrapolation and no majorization, respectively.

Let us examine the algorithmic details.
We choose
\beq \label{eq:majorantG}
G_\lambda(\bx|\bar{\bx}) = f(\bx) - 2 \lambda \langle \bar{\bx}, \bx - \bar{\bx} \rangle - \lambda \| \bar{\bx} \|^2
\eeq
as our convex differentiable majorant
(note that $\| \bx \|^2 \geq \| \bar{\bx} \|^2 + 2 \langle \bar{\bx}, \bx - \bar{\bx} \rangle$, and $f$ is convex differentiable).
The gradient of the above majorant is
\beq \label{eq:nabla_G}
\nabla G_\lambda(\bx| \bar{\bx} ) = - \sum_{i=1}^M \frac{1}{\Phi(\bg_i^T \bx)}
\frac{ e^{-(\bg_i^T \bx)^2/2}}{\sqrt{2\pi}} \bg_i - 2 \lambda \bar{\bx}
\eeq
where
\beq \label{eq:g_i}
\bg_i =\frac{y_i}{\sigma} \bh_i, \quad i=1,\ldots,M.
\eeq
The step size $\beta_t$ is chosen such that the sufficient descent condition
\ifconfver
	\beq \label{eq:suff_des}
	\begin{aligned}
	& G_\lambda(\bx^{t+1}|\bx^t)  \leq
	G_\lambda(\bz^{t}|\bx^t)  \\
	& ~~~ +
	\langle \nabla G_\lambda(\bz^{t}|\bx^t), \bx^{t+1} - \bz^t \rangle
	+ \frac{1}{2 \beta_t} \| \bx^{t+1} - \bz^t \|^2
	\end{aligned}
	\eeq
\else	
	\beq \label{eq:suff_des}
	G_\lambda(\bx^{t+1}|\bx^t) \leq
	G_\lambda(\bz^{t}|\bx^t) +
	\langle \nabla G_\lambda(\bz^{t}|\bx^t), \bx^{t+1} - \bz^t \rangle
	+ \frac{1}{2 \beta_t} \| \bx^{t+1} - \bz^t \|^2
	\eeq
\fi
is satisfied.
We use the backtracking line search \cite{beck2017first} to find such a $\beta_t$.
The extrapolation sequence $\{ \alpha_t \}$ is chosen as that of FISTA~\cite{beck2017first}, given by
\begin{equation}\label{eq:extro_para}
\alpha_t = \frac{\xi_{t-1} - 1}{\xi_t}, \quad \xi_t = \frac{1 + \sqrt{1+4 \xi_{t-1}^2 }}{2},
\quad \xi_{-1}= 1.
\end{equation}

\subsection{A Numerical Issue and Its Fix}
\label{sect:num_fix}

We should mention a numerical issue with the gradient-based algorithm in the preceding subsection.
In the gradient expression in \eqref{eq:nabla_G}, observe that the term $e^{-x^2/2}$ vanishes as $|x|$ is very large.
We found that the occurrence of such vanishing instances can substantially slow down the convergence speed of the gradient-based algorithm, and it may even lead to numerical instability.
As a practical trick, we get this around by over-estimating the noise variance, specifically, by replacing the original $\sigma$ with $\sigma + \sigma_0$ for some fixed $\sigma_0 > 0$.
We found that this trick works empirically.
On the other hand, doing so means that we actually implement a mismatched ML detector
\begin{equation}\label{eq:ML_mis}
\min_{\bx \in \{ -1, 1 \}^N }
-
\sum_{i=1}^M \log \Phi \left(  \frac{ y_i \bh_i^T \bx }{ \hat{\sigma} } \right),
\end{equation}
for some mismatched noise variance $\hat{\sigma}^2$.
By our numerical experience, we do not see noticeable performance degradation with the mismatched ML detector.
We justify this by the following fact.
\begin{Fact} \label{fact:2}
	Consider the same settings as in Fact~\ref{fact:1}, but with the objective function $f$ replaced by that in \eqref{eq:ML_mis}.
	Suppose that the probability density function of $\bh$ is Gaussian with mean zero and covariance $\rho \bI$ for some $\rho > 0$.
	Then,
	the minimizer of the ``large-$M$'' objective function $\tilde{f}(\bx)$ in \eqref{eq:tf} over $\| \bx \|^2 = N$ is uniquely given by the true binary vector.
	It follows that the mismatched ML problem
	\[
	\min_{\bx \in \{ -1, 1 \}^N } \tilde{f}(\bx)
	\]
	has its solution uniquely given by the true binary vector.
	The above results hold for any $\hat{\sigma}^2 > 0$.
\end{Fact}
Fact \ref{fact:2} is a corollary of Lemma 2 in \cite{choi2015quantized},
which considers the no-mismatch case $\hat{\sigma}^2 = \sigma^2$.
Fact \ref{fact:2} suggests that, if the number of antennas at the BS is very large, the ML detector may be insensitive to the mismatch of the noise variance.

{\em Proof of Fact \ref{fact:2}:} \
First, we recall a basic result in stochastic orders \cite{shaked2007stochastic}.
Let $\xi, \upsilon$ be two continuous random variables,
and let
$\phi: \Rbb \rightarrow \Rbb$ be a strictly increasing function.
If
\begin{align} \label{eq:stoc_ord}
{\rm Prob}\{ \xi \geq t \} & \leq {\rm Prob}\{ \upsilon \geq t \}, \quad \text{for all $t$,}
\end{align}
$\xi$ and $\upsilon$ have unequal probability distributions,
then we have
$\Exp[ \phi(\xi) ] >  \Exp[ \phi(\upsilon) ].$

Second, we apply the above result to the problem at hand.
Re-denote the true binary vector in the signal model \eqref{eq:model} as $\bx_0$.
Let $\bx \in \Rbb^N$ be any vector such that $\| \bx \|^2 = N$ and $\bx \neq \bx_0$, and let
\[
\xi = \bh^T \bx_0, \quad \upsilon = \bh^T \bx,
\quad \phi(t) = \log\Phi(t/\hat{\sigma}).
\]
Note that $\phi$ is strictly increasing.
It was shown in Lemma 2 in \cite{choi2015quantized} that \eqref{eq:stoc_ord} holds, and $\xi$ and $\upsilon$ have unequal probability distributions.
Consequently we have
\[
-\tilde{f}(\bx_0) = \Exp[ \phi( \bh^T \bx_0 ) ] > \Exp[ \phi( \bh^T {\bx} ) ] = -\tilde{f}(\bx),
\]
and the minimum of $\tilde{f}(\bx)$ over $\| \bx \|^2 = N$ is attained at, and only at, $\bx_0$.
The proof is complete.
\hfill $\blacksquare$

\section{Making the Homotopy Algorithm a Deep Learnt One}
\label{sect:deep}

As mentioned in the Introduction, lately there has been growing interest in deep learning for MIMO detection via the deep unfolding approach \cite{gregor2010learning,samuel2017deep,Samuel2019learning}.
To employ deep unfolding, one first needs to start with an existing algorithm that is iterative by nature.
The deep unfolding approach sees each iteration of the algorithm as a network layer,
and it seeks to find a better network by untying some of the existing algorithm's parameters and learning those parameters from data.
Additionally, or alternatively, one may make the structure of each iteration more general, with the new structure controlled by some parameters;
and then we learn those parameters from data.

\begin{figure}[h]
	\centering
	\begin{subfigure}[b]{.49\textwidth}
		\includegraphics[width=\textwidth]{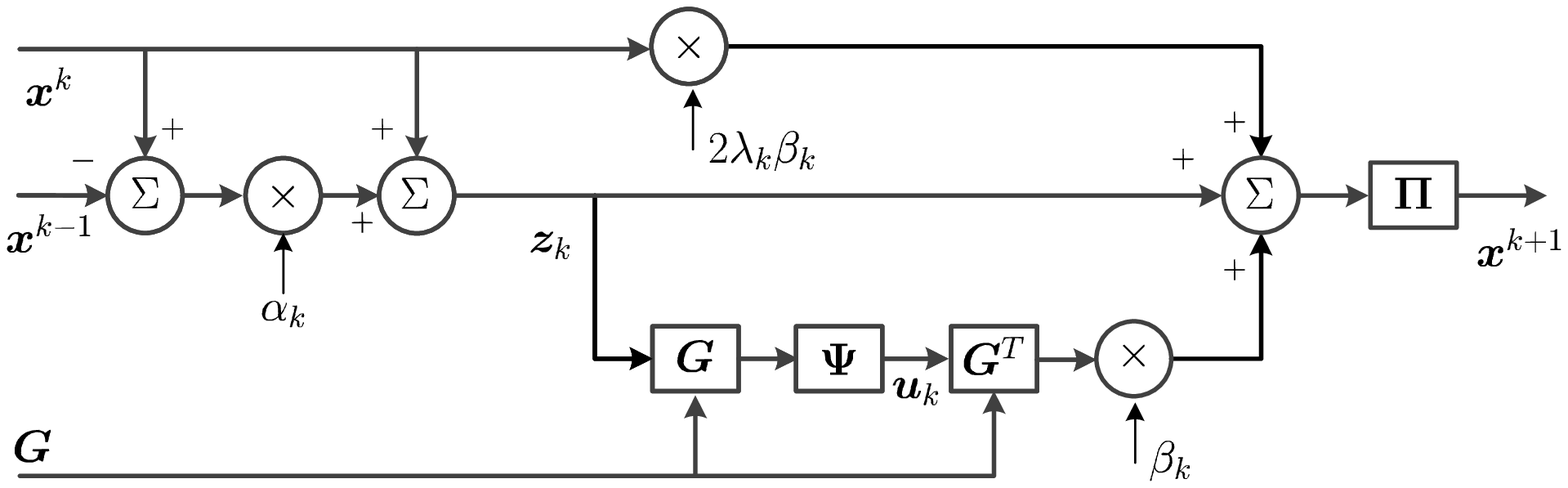}
		\caption{before deep unfolding}
	\end{subfigure}
	\hfill
	\begin{subfigure}[b]{.49\textwidth}
		\includegraphics[width=\textwidth]{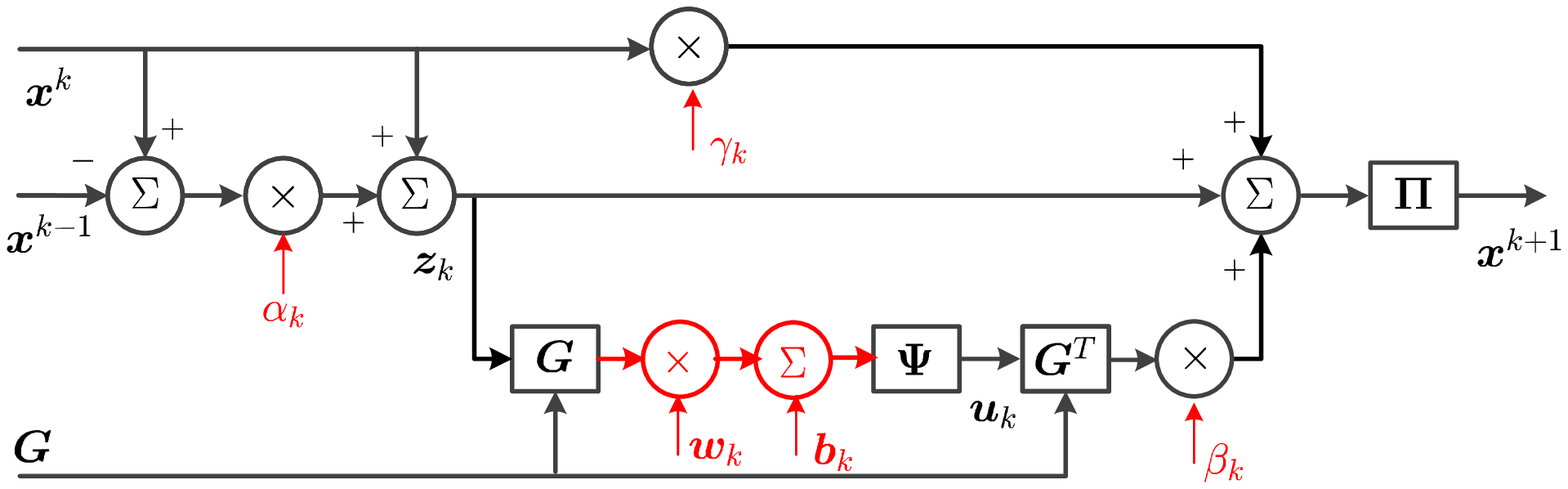}
		\caption{after deep unfolding}
	\end{subfigure}
	\caption{Illustration of the network structure.}\label{fig:DU}
\end{figure}

The homotopy algorithm in the preceding section is very suitable for deep unfolding.
To put into context, we first note that the homotopy algorithm in Algorithm 1, with the descent-based algorithm given by the one in Section~\ref{sect:gemm}, contains two loops---one for the update of the penalty parameter $\lambda_k$, another for the iterative process of the descent-based algorithm.
But we can unfold the two loops into one and write
	\begin{align}
	\bx^{k+1} & = \Pi( \bz^k + \beta_k \nabla G_{\lambda_k}(\bz_k | \bx^k )),
	\label{eq:b4unfold_a}
	\end{align}
for some appropriate $\alpha_k, \beta_k, \lambda_k$; cf. Algorithm 1 and \eqref{eq:gemm}.
By putting the derivation of $\nabla G_\lambda(\bx| \bar{\bx} )$ in \eqref{eq:nabla_G} into \eqref{eq:b4unfold_a}, and recalling the expression of $\bz^k$,
we can represent the whole algorithm by
\begin{subequations} \label{eq:b4unfold2}
	\begin{align}
	\bx^{k+1} & = \Pi( \bz^k + \beta_k \bG^T \bu^k + 2 \beta_k \lambda_k \bx^k ),
	\label{eq:b4unfold2_a} \\
	\bu^k & = \Psi( \bG \bz^k ),
	\label{eq:b4unfold2_b}  \\
	\bz^k & = \bx^k + \alpha_k (\bx^k - \bx^{k-1}),
	\label{eq:b4unfold2_c}
	\end{align}
\end{subequations}
where $\bG = [~ \bg_1, \ldots, \bg_M ~]^T$; $\bg_i$ is defined in \eqref{eq:g_i};
$\Psi(\bx) = [~ \Psi(x_1), \ldots, \Psi(x_N) ~]^T$,
\begin{equation*}
\Psi(x) = \frac{1}{\Phi(x)} \frac{1}{\sqrt{2\pi}} e^{-x^2/2}.
\end{equation*}
As mentioned, deep unfolding sees each iteration in \eqref{eq:b4unfold2} as a network layer, and the whole iterative process a deep network.
Fig.~\ref{fig:DU}(a) illustrates \eqref{eq:b4unfold2} in a network form.
We see that $\Pi$ and $\Psi$ appear as nonlinear activation functions; $\bG$, 
$\alpha_k, \beta_k, \lambda_k$ serve as weights.

We apply deep unfolding to \eqref{eq:b4unfold2} by untying $\alpha_k, \beta_k, \lambda_k$, which are the extrapolation coefficient, step size, and penalty parameter, respectively.
This means that we want to learn, from data, how the penalty parameters are adjusted.
We should recall that the challenge with the homotopy method lies in the selection of the penalty parameters,
and now we use data-driven learning to tackle the challenge.
Similarly, we use data-driven learning to decide the step sizes and extrapolation.
In addition, we make the structure more general by adding element-wise weights and biases in \eqref{eq:b4unfold2_b}.
The network structure arising from the aforementioned untying and modification is given by
\begin{subequations} \label{eq:aftunfold}
	\begin{align}
	\bx^{k+1} & = \Pi( \bz^k + \beta_k \bG^T \bu^k + \gamma_k \bx^k ),
	\label{eq:aftunfold_a} \\
	\bu^k & = \Psi( \bw_k \odot (\bG \bz^k) + \bb_k ),
	\label{eq:aftunfold_b}  \\
	\bz^k & = \bx^k + \alpha_k (\bx^k - \bx^{k-1}),
	\label{eq:aftunfold_c}
	\end{align}
\end{subequations}
where $\gamma_k$ is the untied parameter of $2\beta_k \lambda_k$; $\bw_k, \bb_k \in \Rbb^M$ are a weight and bias, respectively; $\odot$ denotes the element-wise, or Hadamard, product.
Fig.~\ref{fig:DU}(b) illustrates the network structure of \eqref{eq:aftunfold}.
The set of parameters to learn at each layer is $\btheta_k \triangleq \{ \alpha_k, \beta_k, \gamma_k, \bw_k, \bb_k \}$.

In the above deep-unfolded homotopy algorithm, we do not give it a staring point $\bx^0$.
Instead we learn a starting point.
We generate $\bx^0$ by
\beq \label{eq:layer0}
\bx^0 = \Pi( \bW_0 \by + \bb_0 ),
\eeq
where $\bW_0 \in \Rbb^{N \times M}$ is a weight matrix and $\bb_0 \in \Rbb^N$ is a bias; both are to be learnt.
We let $\btheta_0 = \{ \bW_0, \bb_0 \}$.

Our training process is rather standard.
The generative model $\by = \sgn(\bH \bx + \bv )$ in
\eqref{eq:model}, or more precisely, the complex model~\eqref{eq:model_complex} and the subsequent complex-to-real conversion~\eqref{eq:c2r}, is used to generate a large number of training samples $(\bx_i,\bH_i,\by_i,\sigma_i)_{i=1}^R$; note that $R$ is the number of samples,
the $\bx_i$'s are uniform i.i.d. generated, and the $\bH_i$'s are generated from a statistical channel model such as the Rayleigh.
The network parameters $\btheta$'s are obtained by applying stochastic gradient descent to the empirical risk
\begin{equation}\label{eq:loss_train}
 \sum_{i=1}^R \| \bx_i - \varphi(\by_i,\bH_i, \sigma_i; \bm\Theta)  \|^2,
\end{equation}
where the function $\varphi(\by,\bH, \sigma; \bm\Theta)$ represents the input-output relationship of the deep network, parameterized by
$\bm\Theta = (\btheta_k)_{k=0}^K$;
$K$ is the number of layers.

	\begin{Remark}
		We should also describe the big-O complexity of the deep-unfolded homotopy algorithm in \eqref{eq:aftunfold}.
		In addition to the usual operations of floating-point $+,-,x,/,\log,\exp$, etc.,
		the algorithm also requires us to compute $\Phi$;
		recall that $\Phi(t) = \int_{-\infty}^t \frac{1}{\sqrt{2\pi}} e^{-\tau^2/2} d\tau$ is the cumulative distribution of a standard Gaussian distribution.
		The computation of $\Phi$ takes definitely more than one floating-point operations,
		although there exist highly specialized algorithms (e.g., \texttt{erfc} in MATLAB) for efficient computation of $\Phi$.
		In real-world implementations, one may build a dedicated  operation for $\Phi$ to efficiently implement the algorithm in \eqref{eq:aftunfold}.
		It can be verified that the per-iteration complexity of the algorithm in \eqref{eq:aftunfold} is ${\cal O}(M)$ for calling $\Phi$, and ${\cal O}(MN)$ for the usual operations.
		Moreover, the non-deep counterpart of \eqref{eq:aftunfold} has the same big-O per-iteration complexity.
	\end{Remark}

\section{Variation for Classical MIMO Detection}
\label{sect:classical}

The preceding concepts for one-bit MIMO detection can also be applied to classical MIMO detection.
The latter considers the unquantized model
\[ \by = \bH \bx + \bv,
\]
and the corresponding ML detector
\begin{equation}\label{eq:ML_classical}
\hat{\bx}_{\rm ML} = \arg \min_{ \bx \in \{ -1, 1 \}^N } ~ \tfrac{1}{2} \| \by - \bH \bx \|^2.
\end{equation}
The homotopy algorithm in Section \ref{sect:homotopy} directly applies by changing $f$ to $ \tfrac{1}{2} \| \by - \bH \bx \|^2$.
Recalling the algorithm representation in \eqref{eq:b4unfold_a} and \eqref{eq:aftunfold_c},
the homotopy algorithm for the classical MIMO case has the form
\begin{subequations} \label{eq:b4unfold_c}
	\begin{align}
	\ifconfver \hspace*{-.05em} \fi
	\bx^{k+1} & = \Pi( \bz^k -  \beta_k \bH^T \bH \bz^k  + \beta_k \bH^T \by + 2 \beta_k \lambda_k \bx^k ),
	\label{eq:b4unfold-c_a} \\
	\bz^k & = \bx^k + \alpha_k (\bx^k - \bx^{k-1});
	\label{eq:b4unfold_c_b}
	\end{align}
\end{subequations}
(we obtain \eqref{eq:b4unfold-c_a} by applying $\nabla f(\bx) = \bH^T \bH \bx - \bH^T \by$ to \eqref{eq:majorantG} and \eqref{eq:b4unfold_a}).
The step size $\beta_k$ can be chosen as $\beta_k = 1/\| \bH\|_2^2$ where $\| \cdot \|_2$ denotes the spectral norm;
this is a standard method for fulfilling the sufficient descent condition \eqref{eq:suff_des} when $f$ is convex quadratic \cite{beck2017first}.
For the deep-unfolded homotopy algorithm, we modify \eqref{eq:b4unfold_c} as
\begin{subequations} \label{eq:aft_unfold_c}
	\begin{align}
	\bx^{k+1} & = \Pi( \bz^k -  \beta_k \bH^T \bH \bz^k  + \omega_k \bH^T \by + \gamma_k \bx^k ),
	\label{eq:aft_unfold-c_a} \\
	\bz^k & = \bx^k + \alpha_k (\bx^k - \bx^{k-1}),
	\label{eq:aft_unfold_c_b}
	\end{align}
\end{subequations}
where $\alpha_k, \beta_k, \omega_k, \gamma_k$ are the parameters to learn.
Note that we train only four parameters at each layer (without counting the zeroth layer in \eqref{eq:layer0}).
The remaining details are essentially identical.
 The per-iteration complexity of the algorithm in \eqref{eq:aft_unfold_c} is ${\cal O}(N^2)$.

\section{Simulation Results}

\ifconfver
\begin{table*}[htb!]
    \else
\begin{table}[htb!]
\fi
    \centering
    \captionsetup{justification=centering}
    \caption{Summary of tested algorithms.}\label{tb:algo}
    \renewcommand{\arraystretch}{1.2}
    \resizebox{\linewidth}{!}{%
        \begin{tabular}{M{25mm}|M{80mm}| M{20mm} |M{50mm}   }
            \hline
            name  & algorithm, reference & $1$-bit or classical & formulation, approach \\ \hline\hline
            HOTML & homotopy algorithm in Sections~\ref{sect:homotopy} and \ref{sect:classical} & both & ML, homotopy optimization \\ \hline
            DeepHOTML & deep-unfolded homotopy algorithm in Sections~\ref{sect:deep} and \ref{sect:classical} & both & ML, deep unfolding of homotopy opt. \\
            \hline\hline
            ZF &
            zero-forcing detector
            & both & linear\\
            \hline
            nML & near-ML detector~\cite{choi2016near} & 1-bit & ML, convex relaxation \\
            \hline
            nML, two-stage & near-ML detector, then local exhaustive search~\cite{choi2016near} & 1-bit & ML, convex relaxation, local search \\
            \hline
            GAMP & generalized approximate message passing \cite{wang2014multiuser,wen2015bayes} & 1-bit & Bayesian, approximate message passing \\
            \hline \hline
            MMSE DF & minimum-mean-square-error decision-feedback detector   &  classical & linear, decision feedback\\ \hline
            LRA MMSE DF &   lattice reduction-aided MMSE DF detector~ \cite{wubben2011lattice}
            & classical & linear, lattice reduction \\
            \hline
            SD & sphere decoder, Schnorr-Euchner implementation~\cite{damen2003maximum} & classical & ML, exact branch-and-bound search \\
            \hline
            SDR & semidefinite relaxation, implemented by the row-by-row method~\cite{wai2011cheap} & classical & ML, convex relaxation \\
            \hline
            DetNet & detection network \cite{Samuel2019learning} & classical & ML, deep unfolding \\
            \hline
            LAMA & large MIMO approximate message passing~\cite{jeon2018optimal},  damping \cite{Rangan2019on} & classical & Bayesian, approximate message passing \\
            \hline \hline
        \end{tabular}
    }
\ifconfver
    \end{table*}
\else
    \end{table}
\fi

In this section we show an extensive collection of numerical results to examine the performance of the homotopy algorithm and its deep-unfolded adaptation proposed in the preceding sections, and to provide benchmarking with the state-of-the-art algorithms.

\subsection{Simulation and Algorithm Settings}

We first describe how the simulations were prepared.
The procedure is standard:
We randomly generate the signal according to the complex-valued MIMO model \eqref{eq:model_complex} for the one-bit MIMO case,
and its unquantized counterpart $\by_{\rm C}= \bH_{\rm C} \bx_{\rm C} + \bv_{\rm C}$ for the classical MIMO case.
Then, we apply the complex-to-real conversion in \eqref{eq:c2r} to obtain an instance of $(\by, \bH,\bx)$.
Unless otherwise specified, the complex-valued channel $\bH_{\rm C}$ is generated following the element-wise i.i.d. circular Gaussian distribution with mean zero and unit variance (i.e., the Rayleigh channel).
The transmit vector $\bx_{\rm C}$ is generated following the element-wise i.i.d. uniform distribution on the QPSK constellation set $\{ \pm 1 \pm \jj \}$.
We define the SNR as
\[
{\rm SNR} = \Exp[ \| \bH_{\rm C} \bx_{\rm C} \|^2 ] / \Exp[ \| \bv_{\rm C} \|^2 ],
\]
and we use the above formula to determine  the noise variance $\sigma_{\rm C}^2$ (or its real counterpart $\sigma^2= \sigma_{\rm C}^2/2$) when the SNR is given.
We use $100,000$ independently generated instances to evaluate the bit error rates (BERs) of the algorithms under test.
We also evaluate the computational complexities by numerically counting the floating point operations (FLOPs) and also by measuring the actual runtimes on our computer.
For fairness, the evaluations were done on the same platform and on  the same computer; specifically, MATLAB 8.5, and a desktop computer with Intel i7-4770 processor and 16GB memory.

Next, we provide the implementation details with our proposed algorithms.
For convenience, we name the homotopy algorithm in Section~\ref{sect:homotopy} {\em Homotopy OpTimization ML (HOTML)}, and the deep-unfolded homotopy algorithm in Section~\ref{sect:deep}  {\em DeepHOTML};
the same convention applies to their classical MIMO counterparts in Section \ref{sect:classical}.
Let us first consider HOTML based on the representation in Algorithm 1.
We initialize the penalty parameter as $\lambda_0= 0.01$.
We randomly initialize the algorithm by uniformly generating  $\bx^0$ on $[-1,1]^N$.
The update rule of $\lambda_k$ is \eqref{eq:update_lambda}, with $\mu_k =0.1/k$ for the one-bit MIMO case and $\mu_k  = 1/k$ for the classical MIMO case (a standard step-size rule in subgradient methods).
We stop Algorithm 1 when $| \lambda_k - \lambda_{k-1} | \leq 10^{-4}$.
The descent-based algorithm inside Algorithm 1, which is described in Section \ref{sect:gemm}, is stopped when $\| \bx^{t+1} - \bx^t \| \leq 10^{-4}$ or when the number of iterations exceeds a limit; that limit is $300$ for the one-bit MIMO case, and $100$ for the classical MIMO case.
For the remedy of the numerical issue discussed in Section~\ref{sect:num_fix}, we set $\sigma_0=0.5$.

The implementation details with DeepHOTML, which are mostly with training, are as follows.
Unless otherwise specified, the number of network layers is $K= 20$.
We implement the training process by Tensorflow for Python \cite{abadi2016tensorflow}.
The generation of training samples is exactly the same as the signal generation described above.
We use the ADAM stochastic gradient descent optimizer \cite{kingma2014adam}, with exponentially decaying step sizes, for training:
specifically, the initial step size is $0.001$; the step size decays by $0.9$   for the one-bit case, and $0.95$ for the classical MIMO case, once every $500$ training iterations;
each training iteration contains a batch of $500$ training samples;
we train the network with $10,000$ iterations.
In addition, for the one-bit MIMO case, we do the following:
every training sample has its SNR randomly drawn from a range of $5$dB to $22$dB, by uniform distribution;
the network parameters $\bw_i$'s and $\bb_i$'s are initialized as i.i.d. Gaussian random vectors with mean zero and variance $0.01$;
we initialize $\beta_k= 0.01, \gamma_k = 0.001, \alpha_k = 0.5$ for all $k$.
For the classical MIMO case the settings are similar, with the following differences:
the SNR range is changed to $0$dB to $18$dB; we initialize $\beta_k= 0.01, \omega_k = -0.01, \gamma_k =  0.001, \alpha_k = 0.5$ for all $k$.

\begin{figure*}[ht!]
    \centering
    \begin{subfigure}[b]{0.45\textwidth}
        \includegraphics[width=\textwidth]{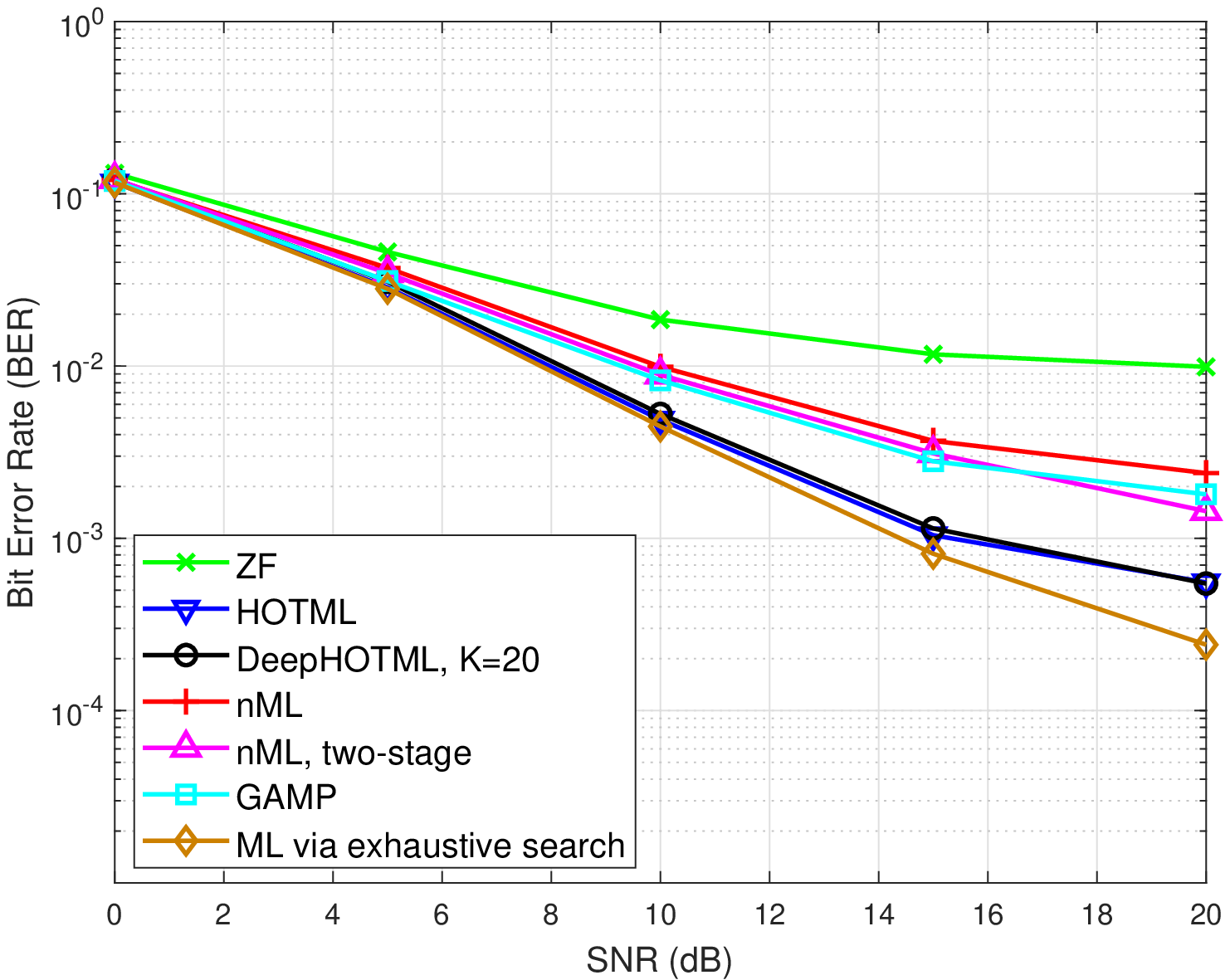}
        \caption{$M= 36$, $N= 8$}
    \end{subfigure}
    ~ 
    \begin{subfigure}[b]{0.45\textwidth}
        \includegraphics[width=\textwidth]{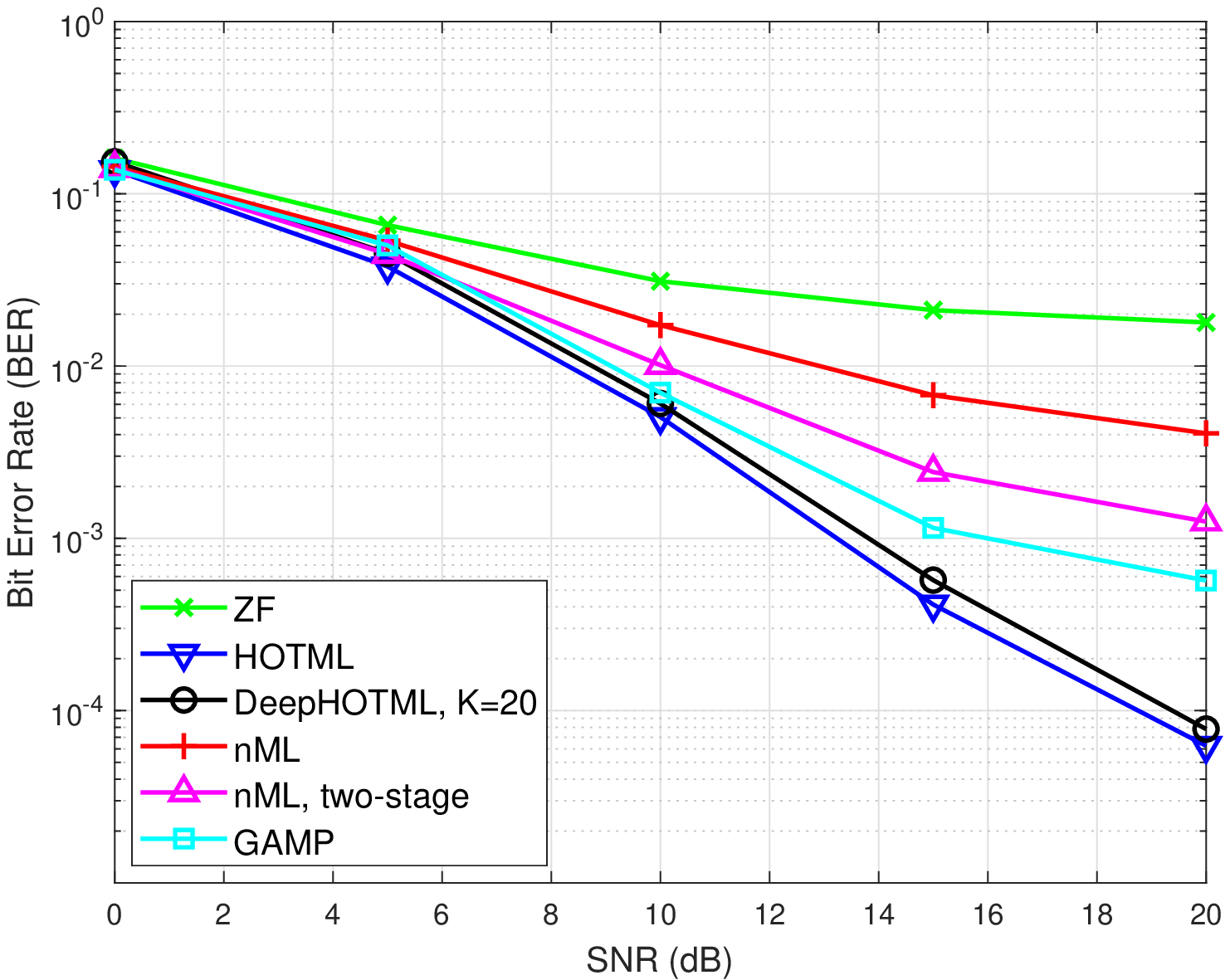}
        \caption{$M= 128$, $N= 32$}
    \end{subfigure}
    ~ 
    \begin{subfigure}[b]{0.45\textwidth}
        \includegraphics[width=\textwidth]{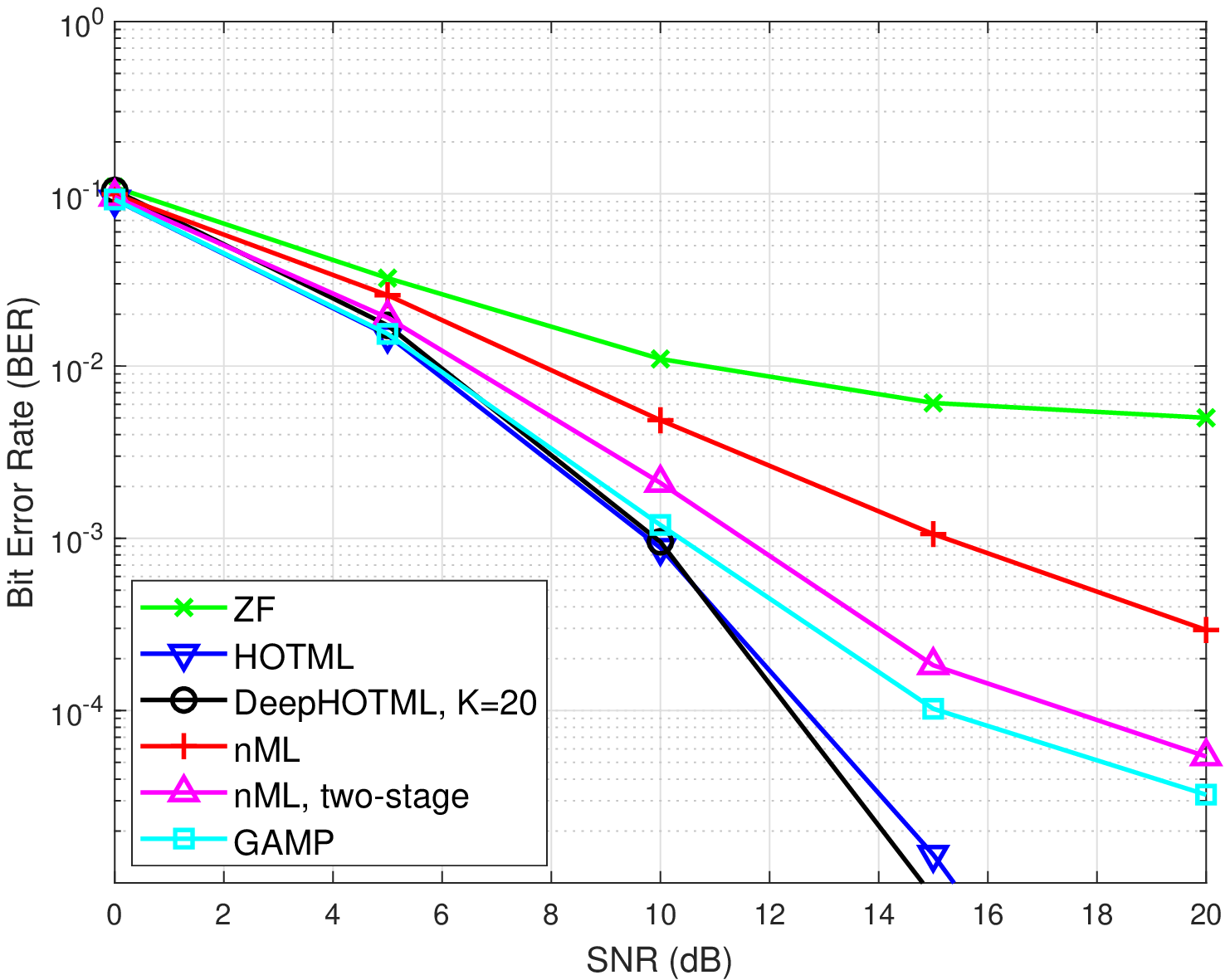}
        \caption{$M= 256$, $N= 48$}
    \end{subfigure}
    ~ 
    \begin{subfigure}[b]{0.45\textwidth}
        \includegraphics[width=\textwidth]{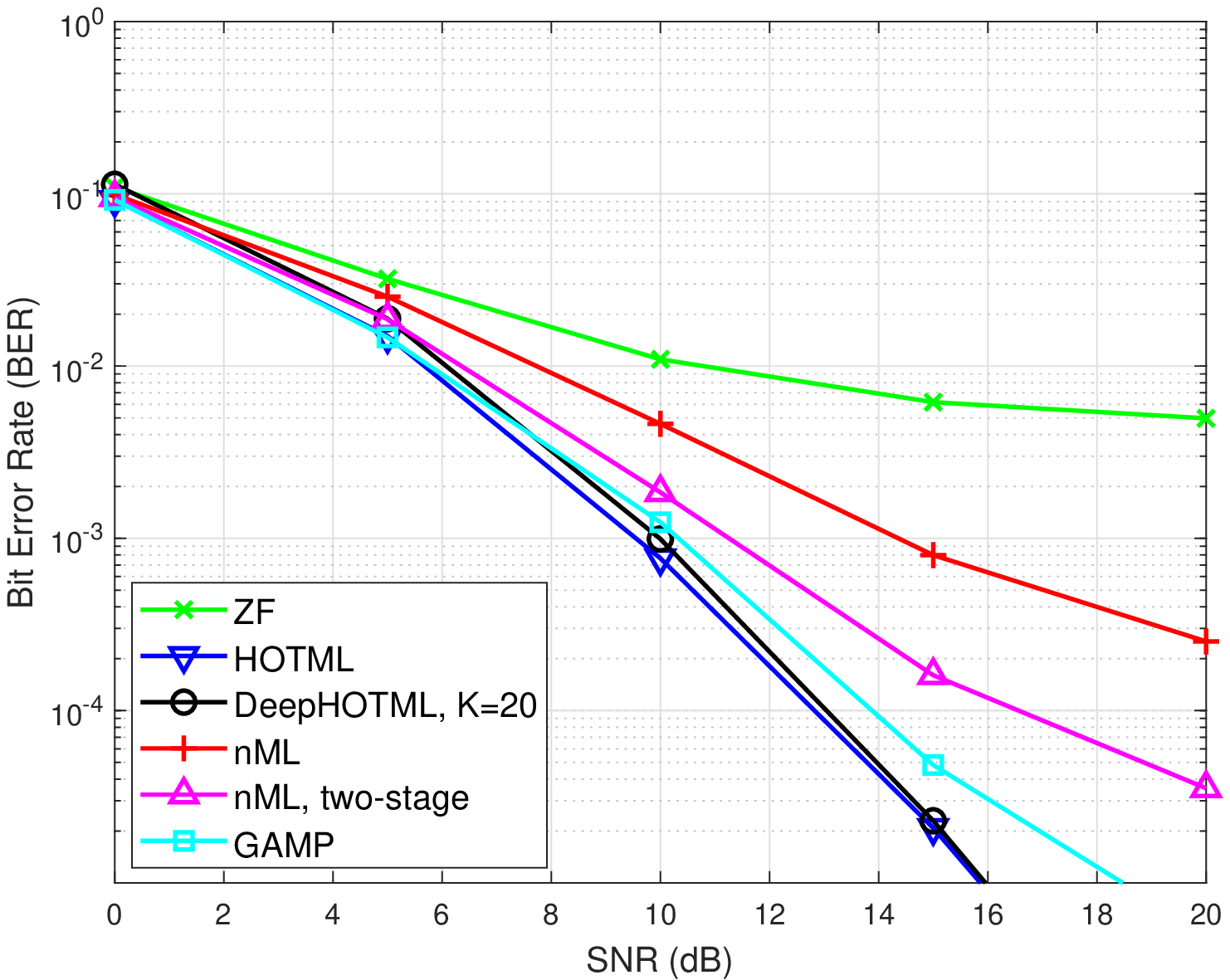}
        \caption{$M= 512$, $N= 96$}
    \end{subfigure}

    \caption{One-bit MIMO detection performance.}\label{fig:ob_ber}
\end{figure*}

The algorithms we benchmarked should also be mentioned.
Table~\ref{tb:algo} provides a summary of the tested algorithms.
These algorithms either are some of the {\em de facto} standards in MIMO detection, or are emerging methods.
Some additional implementation details should be mentioned.
We modify nML~\cite{choi2016near} by changing its step-size rule to the backtracking line search, which is also used in HOTML for one-bit MIMO detection;
we found that this modification improves the performance of nML.
The number of iterations of GAMP is set to $20$, as recommended in \cite{wang2014multiuser}.
The number of iterations of the row-by-row method~\cite{wai2011cheap} for SDR is $200$.
The number of iterations of LAMA is $100$, the same one used in~\cite{jeon2018optimal}; also we stop LAMA if the difference of successive iterates is less than $10^{-4}$.
DetNet~\cite{Samuel2019learning} has its source code available at \url{https://github.com/neevsamuel/LearningToDetect}.
We use that open source code to train DetNet under the recommended settings of $30$ network layers, $50,000$ training iterations, and $3,000$ training samples for each training iteration.

\ifconfver
\begin{table*}[htb!]
    \else
    \begin{table}[htb!]
        \fi
        \centering
        \captionsetup{justification=centering}
        \caption{Complexity comparison with the one-bit MIMO detection algorithms. \\[-0.4em]}\label{tb:flop_onebit}
        \renewcommand{\arraystretch}{1.2}
        \resizebox{0.9\linewidth}{!}{%
            \begin{tabular}{M{20mm}|M{25mm}| M{30mm} M{30mm} M{30mm} M{30mm}  }
                \hline
                $M\times N$ &  & HOTML & DeepHOTML & nML & GAMP \\ \hline\hline
                \multirow{3}{*}{$128\times 32 $} & FLOPs & $1.8862\times 10^7$ & $3.5187\times 10^5$ & $5.6806\times 10^5$ &   $3.9269 \times 10^5$
               \\
                & no. of $\Phi$ calc.  & $1.4555\times 10^5$ & $2.560\times 10^3$ & $4.5408 \times 10^3$ &   $2.560\times 10^3$
                \\
                & time (Sec.)  &0.0589  & 0.0008 & 0.0027 & 0.0013 \\ \hline
                \multirow{3}{*}{$256\times 48$} & FLOPs & $5.1287\times 10^7$  & $1.03801\times 10^6$  & $1.4811\times 10^6$  &  $1.1714\times 10^6$
                \\
                & no. of $\Phi$ calc. & $2.6557\times 10^5$ & $5.120\times 10^3$ & $8.2565\times 10^3$ &   $5.120\times 10^3$
                \\
                & time (Sec.) & 0.0936 & 0.0014 & 0.0053 &  0.0021 \\ \hline
                \multirow{3}{*}{$512 \times 96$ } & FLOPs & $9.8408\times 10^7$ & $4.0913\times 10^6$ & $8.1409\times 10^6$ &   $4.2355\times 10^6$ 
                \\
                & no. of $\Phi$ calc. & $3.8483\times 10^5$ & $1.024\times 10^4 $ & $2.0065\times 10^{4}$ &   $1.024\times 10^4 $
                \\
                & time (Sec.) & $0.1308$ & $0.0019$ & $0.0088$ & $0.0029$ \\
                \hline
            \end{tabular}
        }
        \ifconfver
    \end{table*}
    \else
\end{table}
\fi

\subsection{One-Bit MIMO Detection}

In this subsection we examine the one-bit MIMO case.
Fig.~\ref{fig:ob_ber} shows the BERs of the tested algorithms under different problem sizes $(M,N)$.
 The benchmarked algorithms are
 the ZF detector, which is the  direct application of zero forcing in classic MIMO detection; nML and its two-stage variant, which are based on sphere relaxation of the ML problem \eqref{eq:prob_main} (cf., problem \eqref{eq:sr}); and GAMP which is the application of approximate message passing to one-bit MIMO detection.
For the small problem size case of 	$(M,N) = (36,8)$ in Fig.~\ref{fig:ob_ber}(a), we  also provide a performance baseline by evaluating the performance of the optimal ML detector via exhaustive search.
We see from Fig.~\ref{fig:ob_ber} that HOTML and DeepHOTML generally yield similar BER performance and achieve much better BER performance than all the other benchmarked algorithms except the exhaustive search.
It is also observed that the performance of GAMP improves as the problem size increases.

Next, we compare the computational complexities of the tested algorithms.
As mentioned in Remark 2,
 HOTML and DeepHOTML require calling the function $\Phi(t) = \int_{-\infty}^t \frac{1}{\sqrt{2\pi}} e^{-\tau^2/2} d\tau$.
Hence we divide the operation counts into two, one as the FLOPs of the usual operations ($+$, $-$, $\times$, $/$, $\log$, $\exp$), the other as the number of times  $\Phi$ is called.
Note that nML and GAMP also require calling $\Phi$ \footnote{ GAMP requires computing the expectation and variance of some posterior distribution, and this can be done by computing $\Phi$ \cite{wen2015bayes}.}, and their operational counts are done by the same way as above.

\begin{figure}[ht!]
    \centering
    \begin{subfigure}[b]{0.45\textwidth}
        \includegraphics[width=\textwidth]{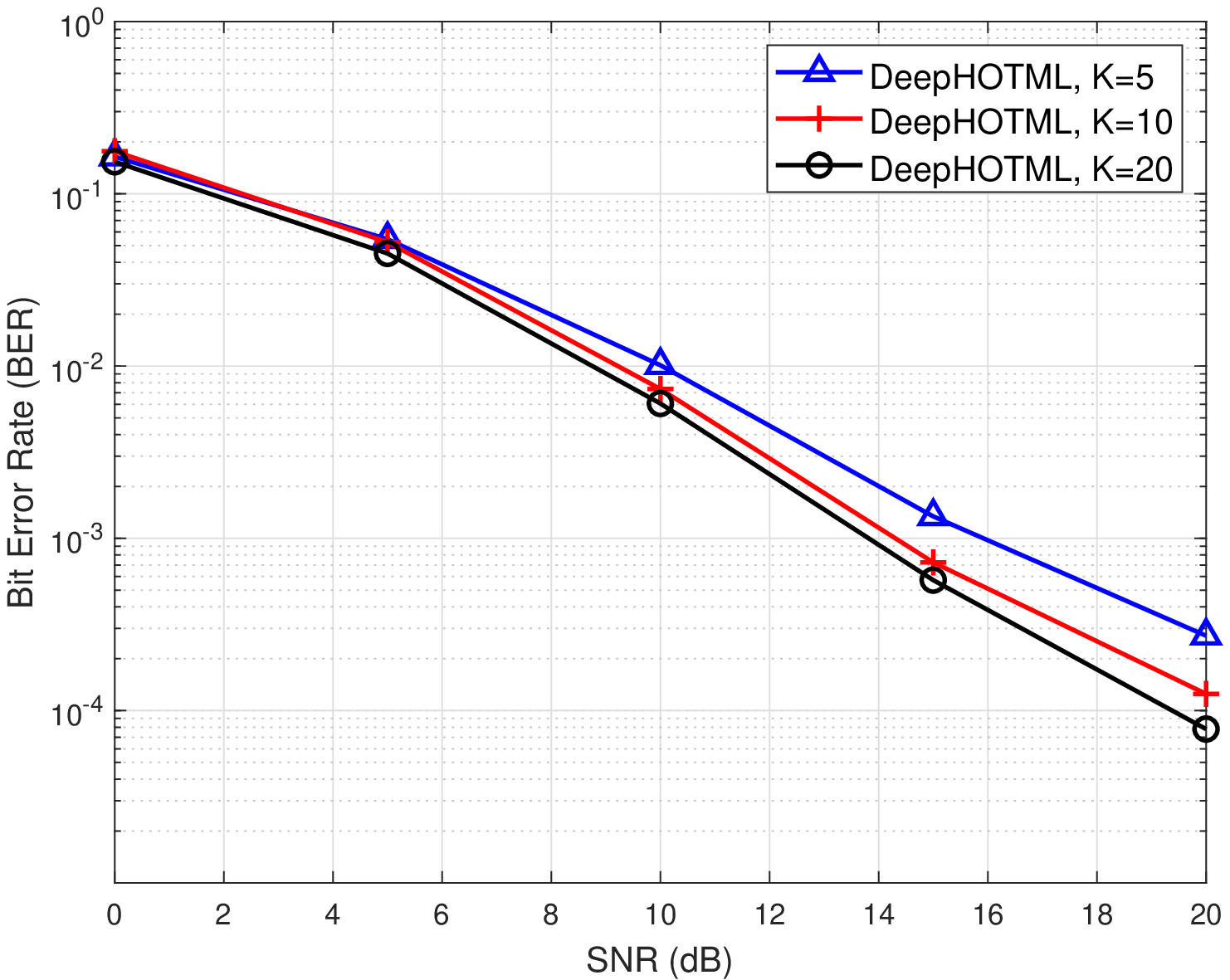}
        \caption{$M= 128$, $N= 32$}
    \end{subfigure}
    ~ 
    \begin{subfigure}[b]{0.45\textwidth}
        \includegraphics[width=\textwidth]{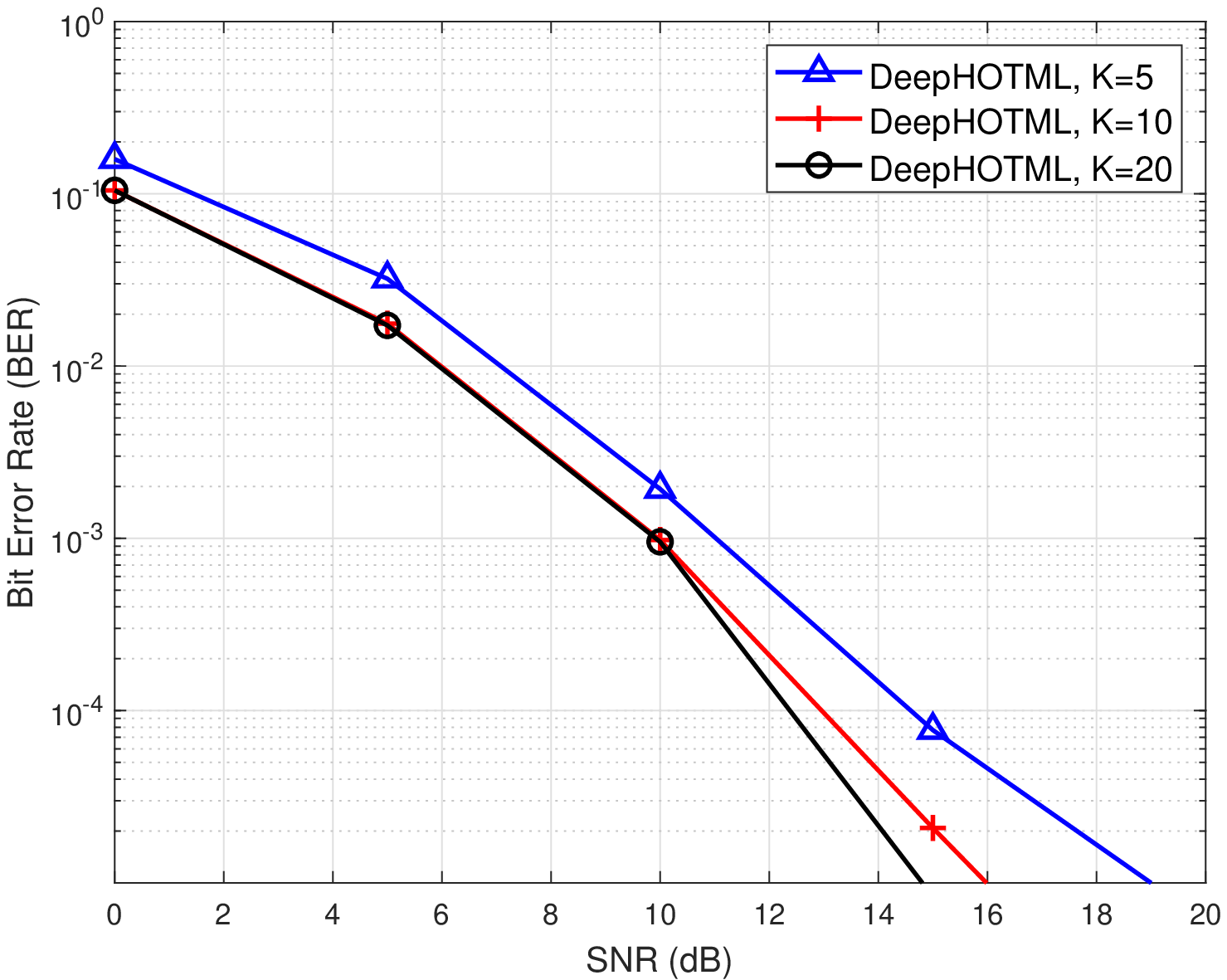}
        \caption{$M= 256$, $N= 48$}
    \end{subfigure}
    ~ 
    \begin{subfigure}[b]{0.45\textwidth}
        \includegraphics[width=\textwidth]{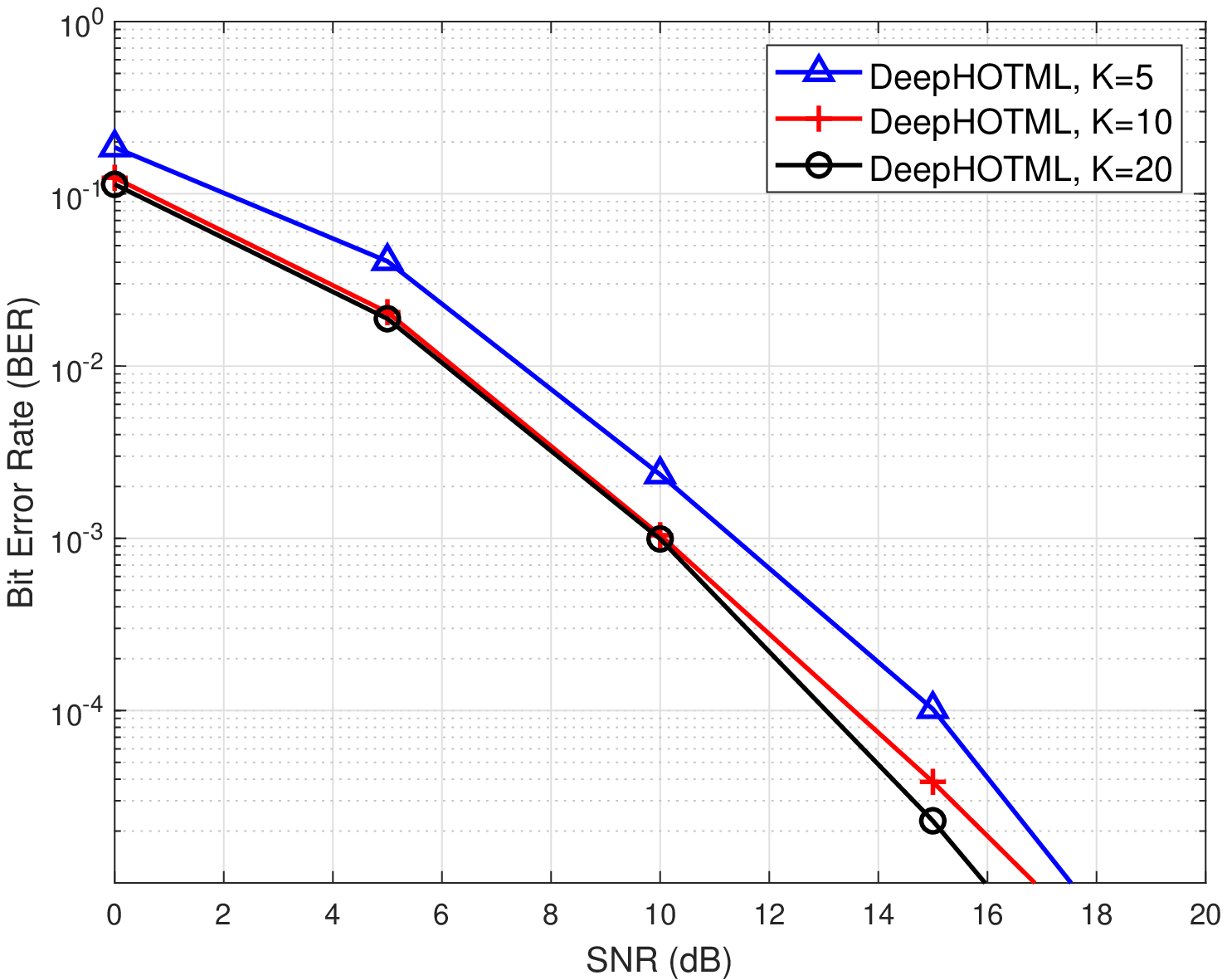}
        \caption{$M= 512$, $N= 96$}
    \end{subfigure}

    \caption{One-bit MIMO detection performance of DeepHOTML under different numbers of layers.}\label{fig:ob_ber_deep}
\end{figure}

The complexity results are shown in Table~\ref{tb:flop_onebit}.
We use the same settings as in Fig.~\ref{fig:ob_ber}, with the SNR fixed at $15$dB.
The fastest algorithm, as revealed by both the actual runtimes and operation counts, is DeepHOTML.
GAMP and nML are the second and third best, respectively; HOTML is the slowest.
It is worth noting that the computations of DeepHOTML are similar to those of a gradient descent method with a fixed number of iterations (or layers) of $20$---this is why DeepHOTML is fast.
The computations of HOTML are also similar to those of a gradient descent method.
However, as suggested by Table~\ref{tb:flop_onebit}, HOTML uses considerably more iterations than DeepHOTML;
in fact,  HOTML's line search routine for determining the step sizes also costs a non-negligible amount of computations.

It is also interesting to examine the BER performance of DeepHOTML under different number of network layers.
Fig.~\ref{fig:ob_ber_deep} shows the results; again, the settings are identical to those in Fig.~\ref{fig:ob_ber}.
Note that when we change the number of layers $K$, we train DeepHOTML for that particular $K$.
We observe from Fig.~\ref{fig:ob_ber_deep} is that the BER performance improves as the number of layers increases.
Also,
it is encouraging to see that DeepHOTML with $10$ layers can achieve BER performance similar to that with $20$ layers; DeepHOTML with 5 layers also  looks good.


\subsection{Classical MIMO Detection}
\begin{figure}[ht!]
    \centering
    \begin{subfigure}[b]{0.45\textwidth}
        \includegraphics[width=\textwidth]{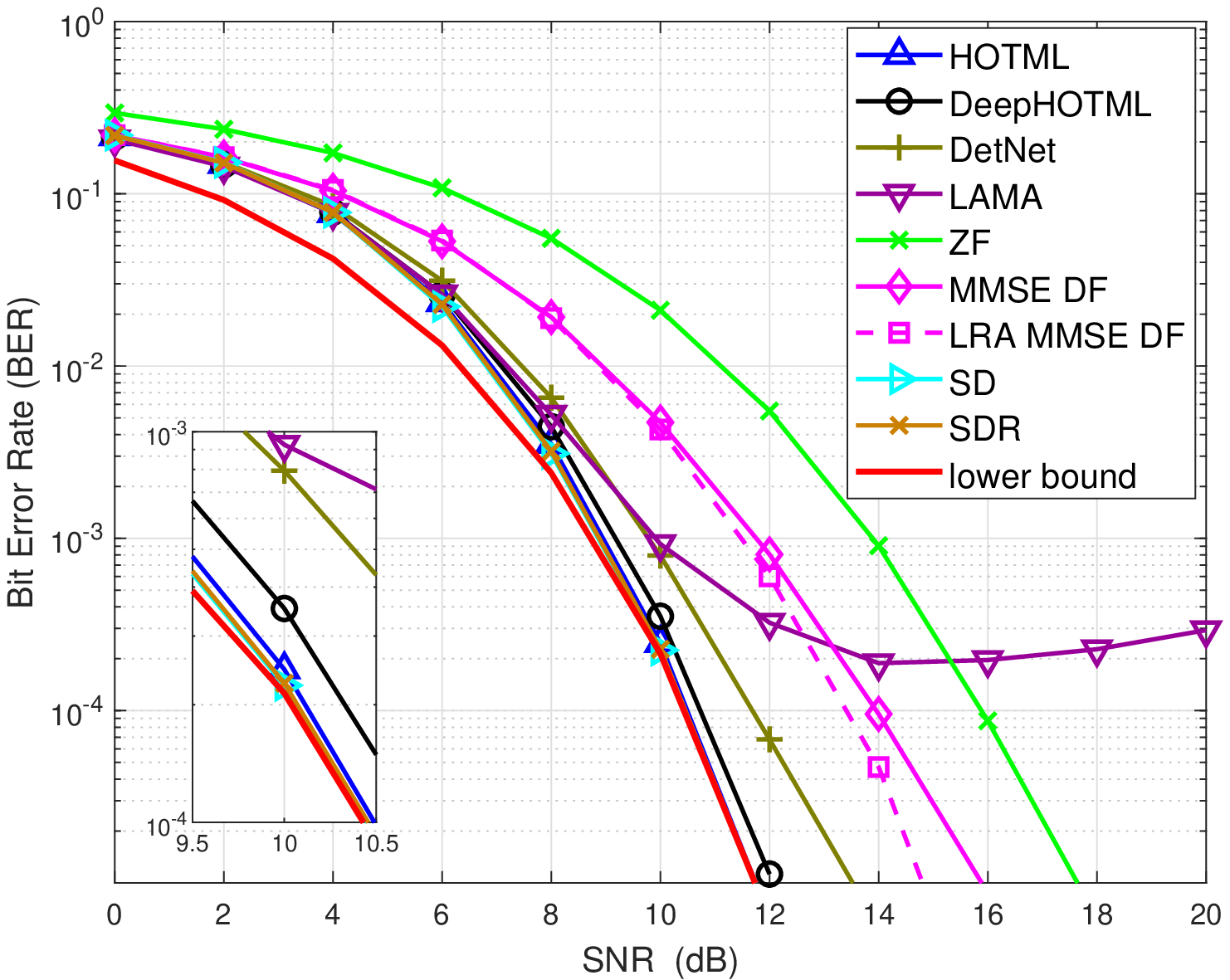}
        \caption{$M= 60$, $N= 40$}
    \end{subfigure}
    ~ 
    \begin{subfigure}[b]{0.45\textwidth}
        \includegraphics[width=\textwidth]{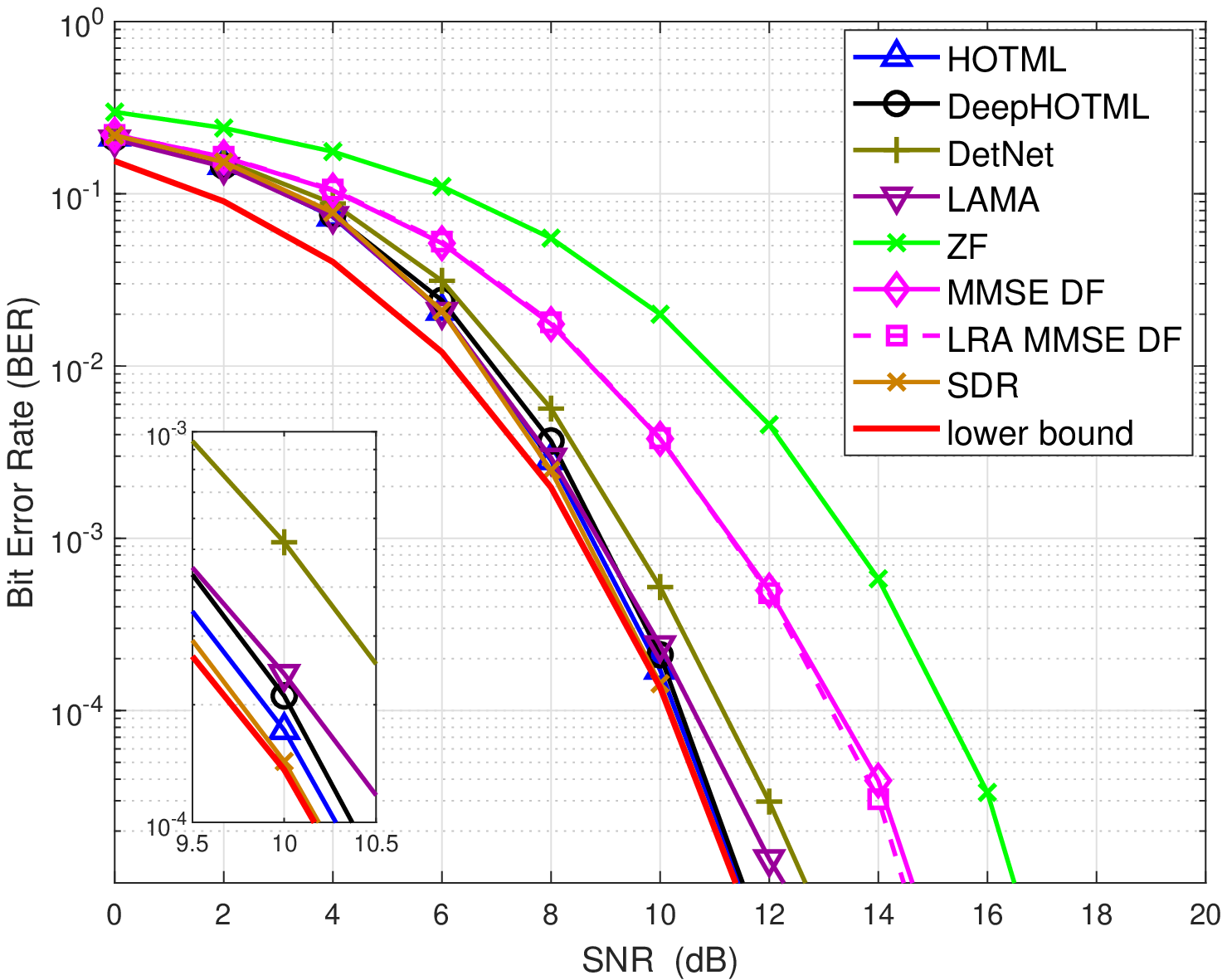}
        \caption{$M= 120$, $N= 80$}
    \end{subfigure}

    \caption{Classical MIMO detection performance for overdetermined channel cases.}\label{fig:class_ber_overdet}
\end{figure}

Now we turn to the classical MIMO case.
Fig.~\ref{fig:class_ber_overdet} shows the BERs of the tested algorithms for the relatively easier instances of overdetermined MIMO channels (i.e., $M > N$).
  In the figures, ``lower bound'' is the BER performance when there is no interference between symbols.
Note that SD is an optimal ML algorithm using some smart branch-and-bound search method, but still it is computationally prohibitive when $N$ is large.
In our simulation, we only tested SD for $N= 40$.
Some observations are as follows.
LAMA suffers from error floor effects for the case of $(M,N)= (60,40)$, but it achieves near-optimal performance when the problem size increases to $(M,N)= (120,80)$.
LAMA appears to be favorable for large MIMO regimes, and we will see the same behaviors later.
Also,
SDR, DetNet, HOTML and DeepHOTML all achieve near-optimal performance.

\begin{figure*}[ht!]
    \centering
    \begin{subfigure}[b]{0.46\textwidth}
        \includegraphics[width=\textwidth]{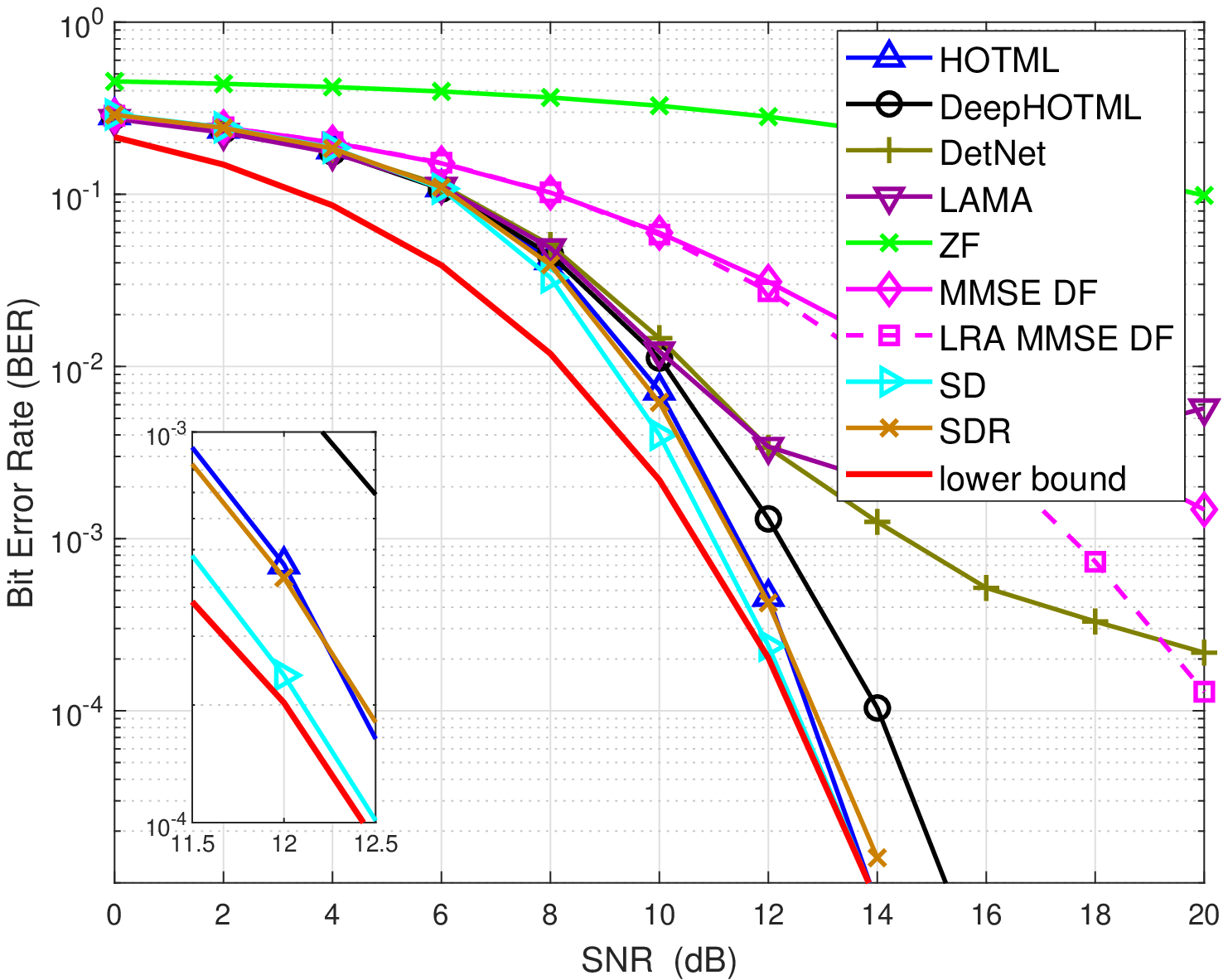}
        \caption{$M= N= 40$}
    \end{subfigure}
    ~
    \begin{subfigure}[b]{0.46\textwidth}
        \includegraphics[width=\textwidth]{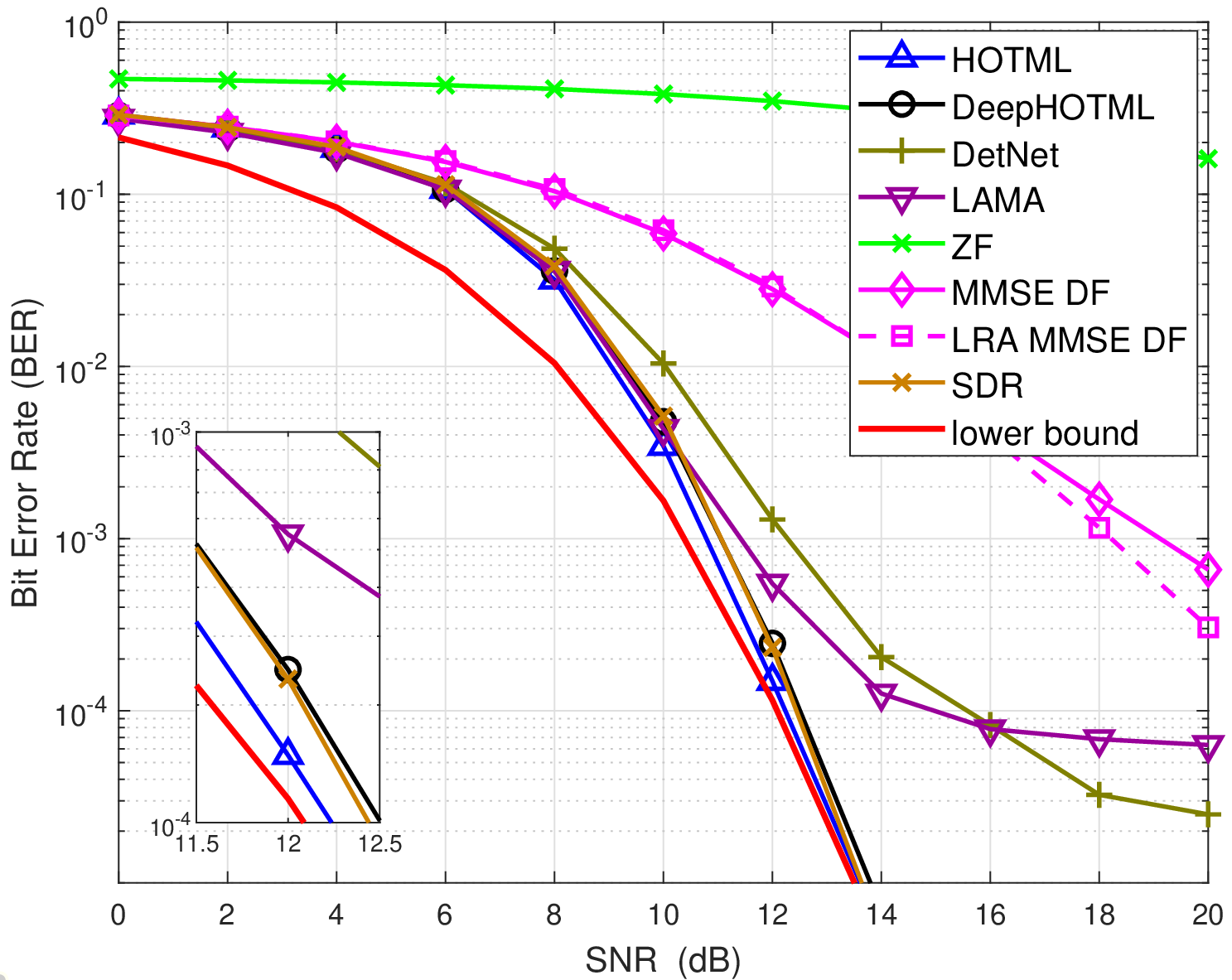}
        \caption{$M= N= 80$}
    \end{subfigure}
    ~ 
    \begin{subfigure}[b]{0.46\textwidth}
        \includegraphics[width=\textwidth]{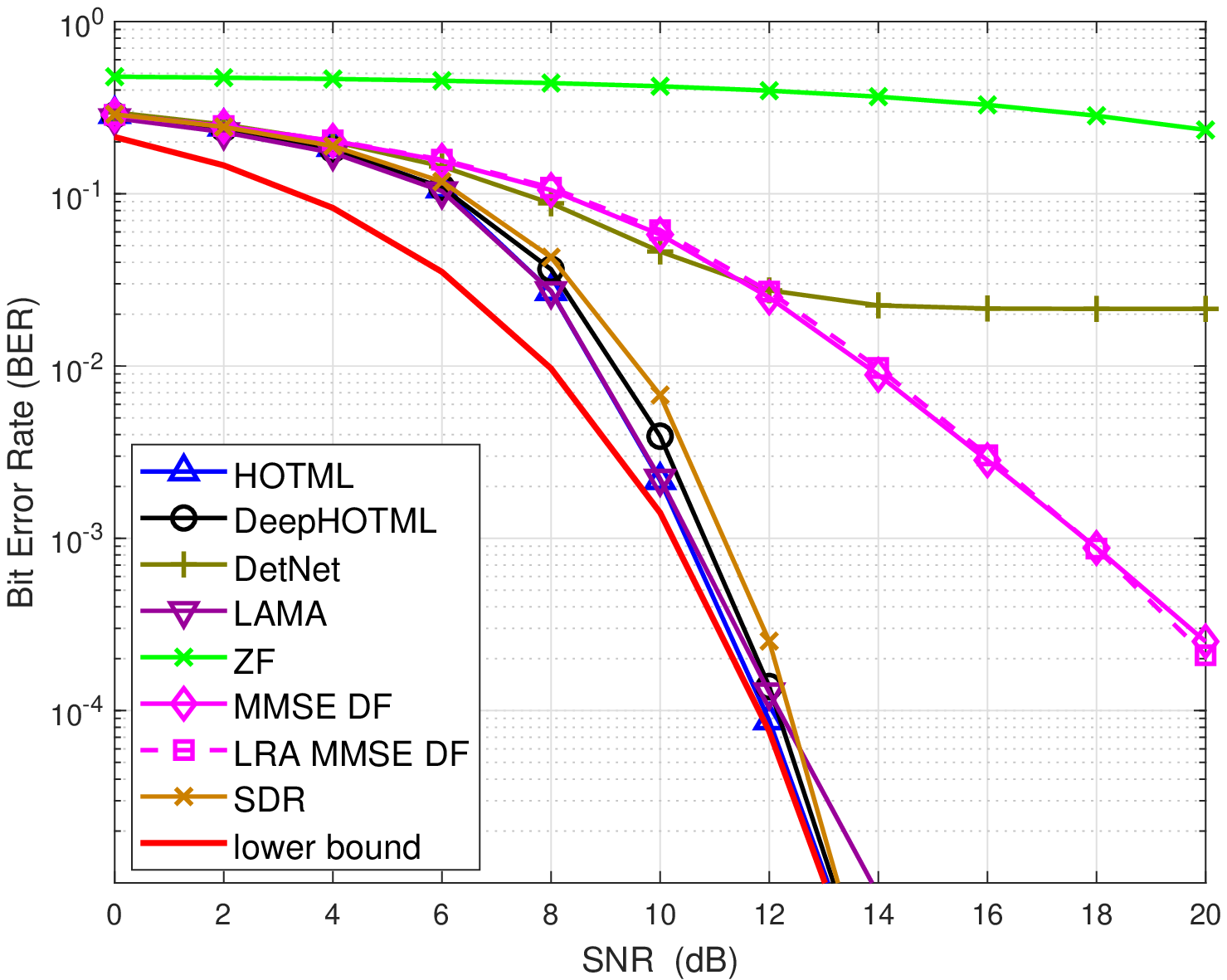}
        \caption{$M= N= 160$}\label{fig:FigBER_iid_80by80}
    \end{subfigure}
    ~ 
    \begin{subfigure}[b]{0.46\textwidth}
        \includegraphics[width=\textwidth]{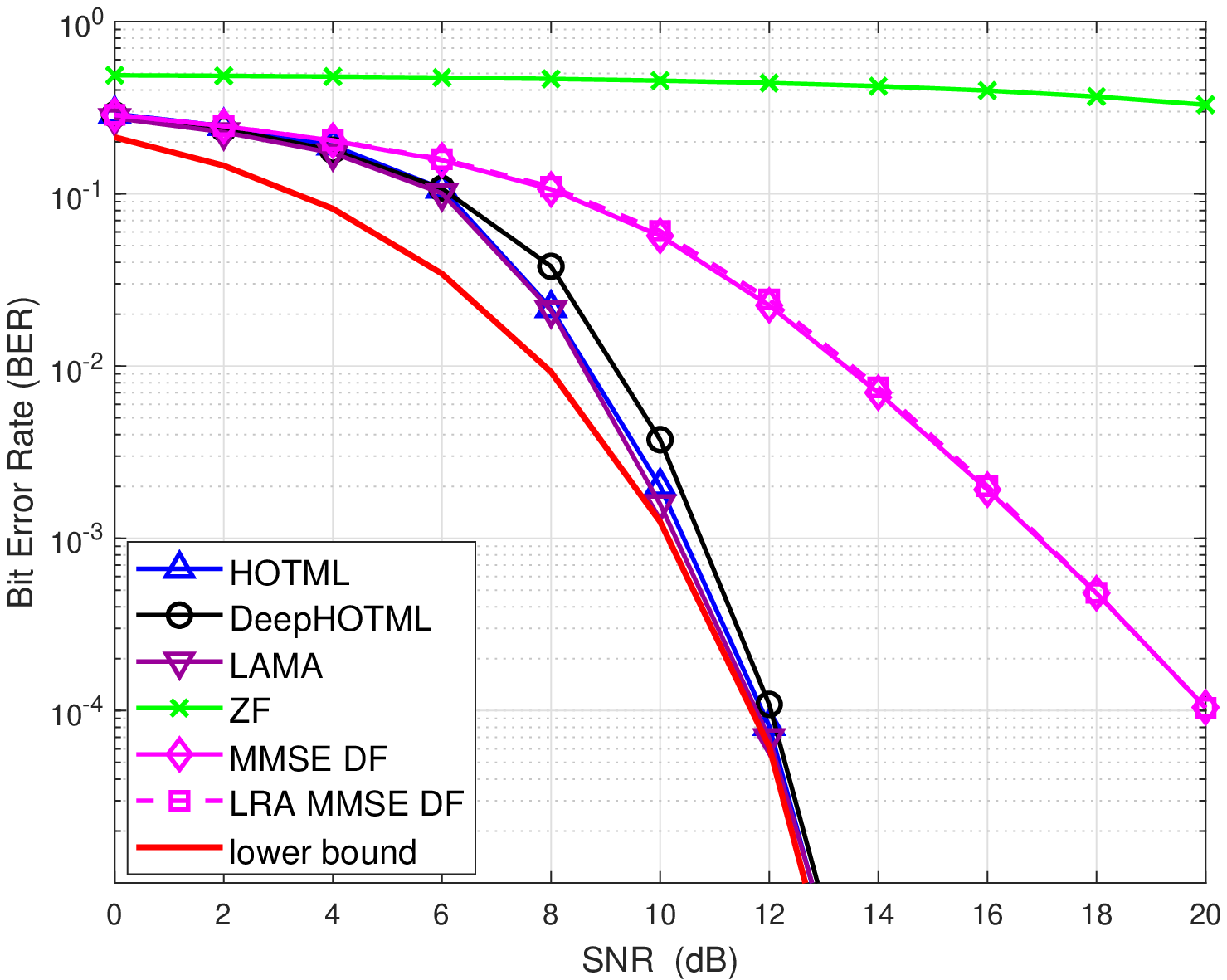}
        \caption{$M= N = 400$}
    \end{subfigure}

    \caption{Classical MIMO detection performance for critically determined channel cases.}\label{fig:class_ber_crit}
\end{figure*}

In the BER plots in Fig.~\ref{fig:class_ber_crit}, we examine the more challenging instances of critically determined channels (i.e., $M= N$).
It is worth noting that DetNet, in its original paper \cite{samuel2017deep,Samuel2019learning}, did not consider the critically determined cases.
We tried our best to train DetNet in critically determined cases, but we were not too successful with obtaining satisfactory results as in the preceding overdetermined cases.
In the case of $M= N= 400$, we were even unable to complete the training; we found that the training (implemented by Tensorflow) drew a very substantial amount of computational resources.
Fortunately, for DeepHOTML, we did not encounter the same problem;
this is likely because our DeepHOTML network has only 4 parameters per layer (without counting the $0$th layer in \eqref{eq:layer0}).
Also, note that we did not run SDR for the case of $M=N= 400$; although SDR is known to have polynomial-time complexity, it is still too expensive to run SDR when the problem size is very large.
We observe that
DetNet does not perform well,
LAMA performs well only when the problem sizes are large,
and SDR, HOTML and DeepHOTML consistently show near-optimal performance;
DeepHOTML performs slightly worse for the small size $M=N=40$ case.

\begin{figure*}[ht!]
    \centering
 \begin{subfigure}[b]{0.46\textwidth}
        \includegraphics[width=\textwidth]{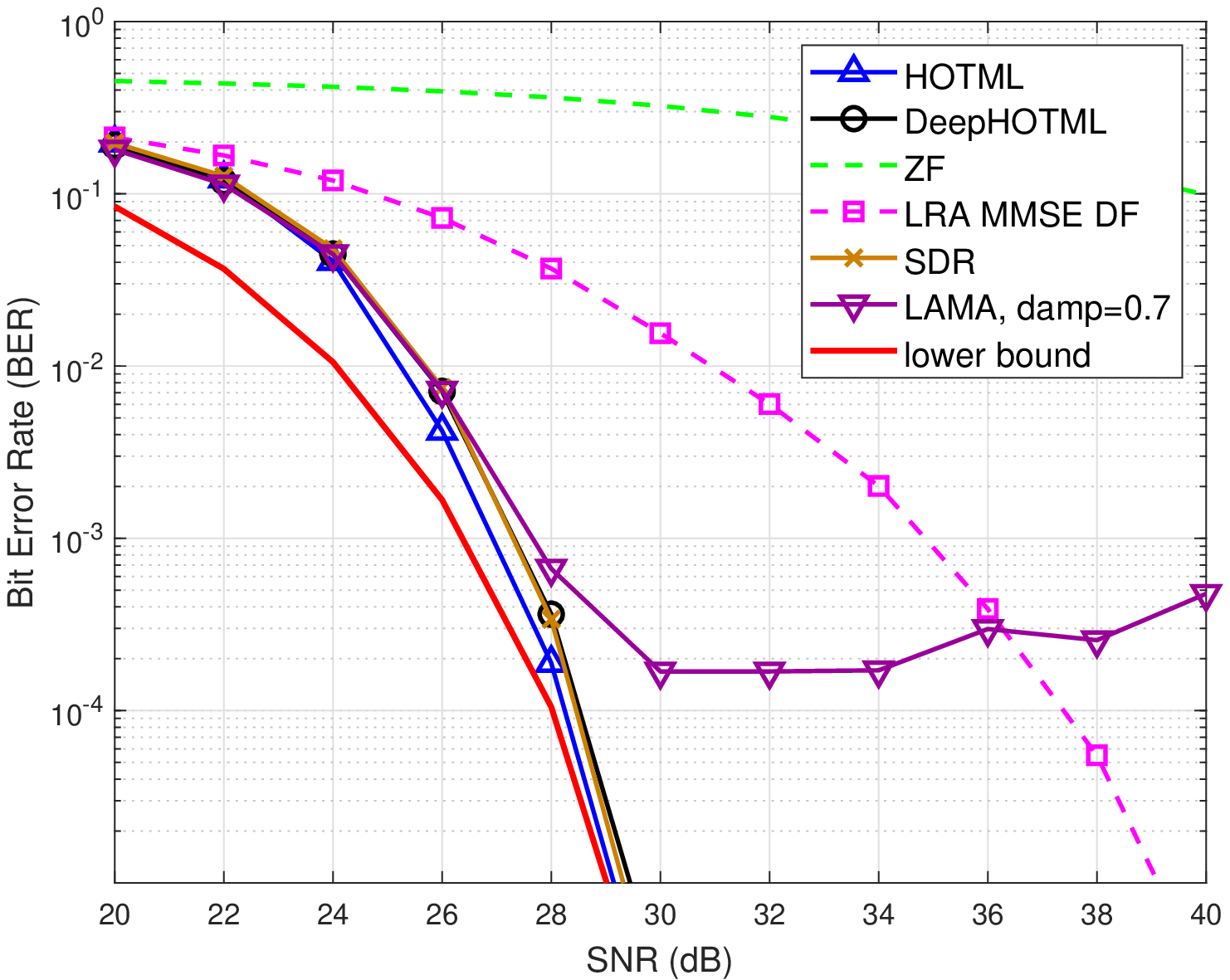}
        \caption{$M= N= 80$}
    \end{subfigure}
    ~
 \begin{subfigure}[b]{0.46\textwidth}
        \includegraphics[width=\textwidth]{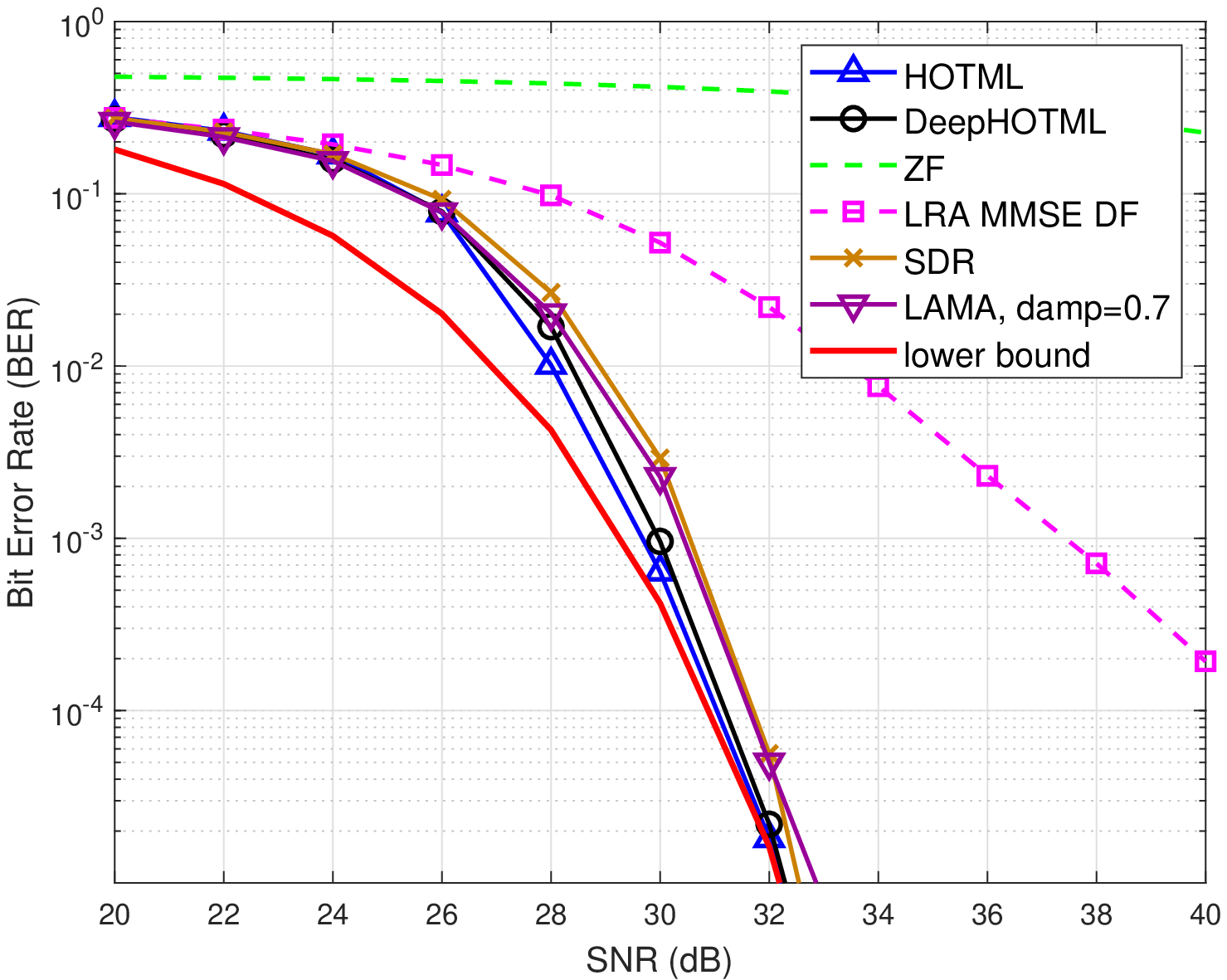}
        \caption{$M= N= 160$}
    \end{subfigure}

    \caption{Classical MIMO detection performance for correlated channels.}\label{fig:BER_corr_80by80}
    \end{figure*}


In addition to the i.i.d. Gaussian channel simulations shown above, we are also interested in correlated MIMO channels, specifically,
\begin{equation*}\label{eq:corr_channel}
  \bH_C= \bR_r^{1/2} \tilde{\bH} \bR_t^{1/2},
\end{equation*}
where
$\tilde{\bH}$ is element-wise i.i.d. circular Gaussian with mean zero and unit variance;
$\bR_{r}$ and $\bR_t$ are the spatial correlation matrices at the receiver and transmitter, respectively, and they are modeled as
\begin{equation*}
  [\bR_r]_{i,j}=\begin{cases}
                  r^{i-j}, & \mbox{if } i\leq j \\
                  [\bR_r]_{i,j}^*, & \mbox{otherwise}
                \end{cases}
\end{equation*}
for $|r|\leq 1$ \cite{loyka2001channel}. In the simulation below, we set $r=0.2$.
Also, DeepHOTML is trained under the SNR range of $18$dB to $34$dB.
The other simulation settings are same as the i.i.d. Gaussian case above.
It is worth noting that LAMA works poorly for correlated channels because it exploits the assumption of i.i.d. Gaussian channels \cite{jeon2018optimal}.
To mend this issue we apply the recently proposed damping technique in approximate message passing \cite{Rangan2019on} to LAMA;
the damping rate is ${\sf damp}=0.7$.
LAMA with damping leads to a slower convergence by our empirical experience (the per-iteration complexity is almost twice of that of the original LAMA), but the performance is improved significantly.
 Fig.~\ref{fig:BER_corr_80by80} shows the BER results for different problem sizes.
It is seen that both HOTML and DeepHOTML achieve near-optimal detection performance; SDR is only slightly worse;
LAMA with damping works well in the large-scale case in Fig.~\ref{fig:BER_corr_80by80}(b), but suffers from error floor effects for the moderate size case  in Fig.~\ref{fig:BER_corr_80by80}(a).

\begin{figure}[ht!]
    \centering
    \includegraphics[width=0.45\textwidth]{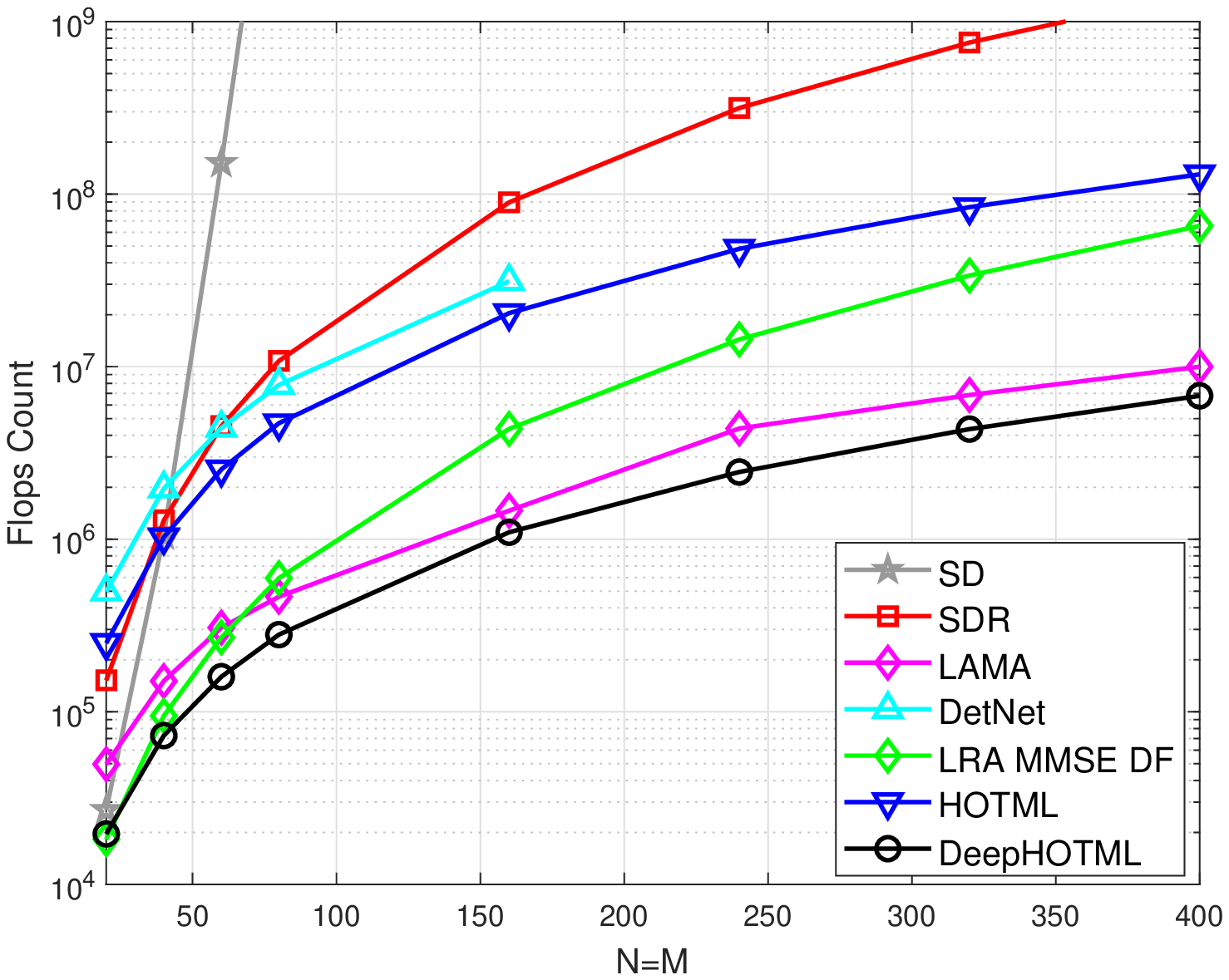}
    \caption{Complexity comparison with the classical MIMO detection algorithms.}\label{fig:flops}
\end{figure}

In Fig.~\ref{fig:flops} we compare the complexities of the tested algorithms by examining their FLOPs under various problem sizes $N$; the SNR is fixed at $10$dB.
We see that DeepHOTML stands as the fastest algorithm in general.

\begin{figure}[ht!]
    \centering
    \begin{subfigure}[b]{0.45\textwidth}
        \includegraphics[width=\textwidth]{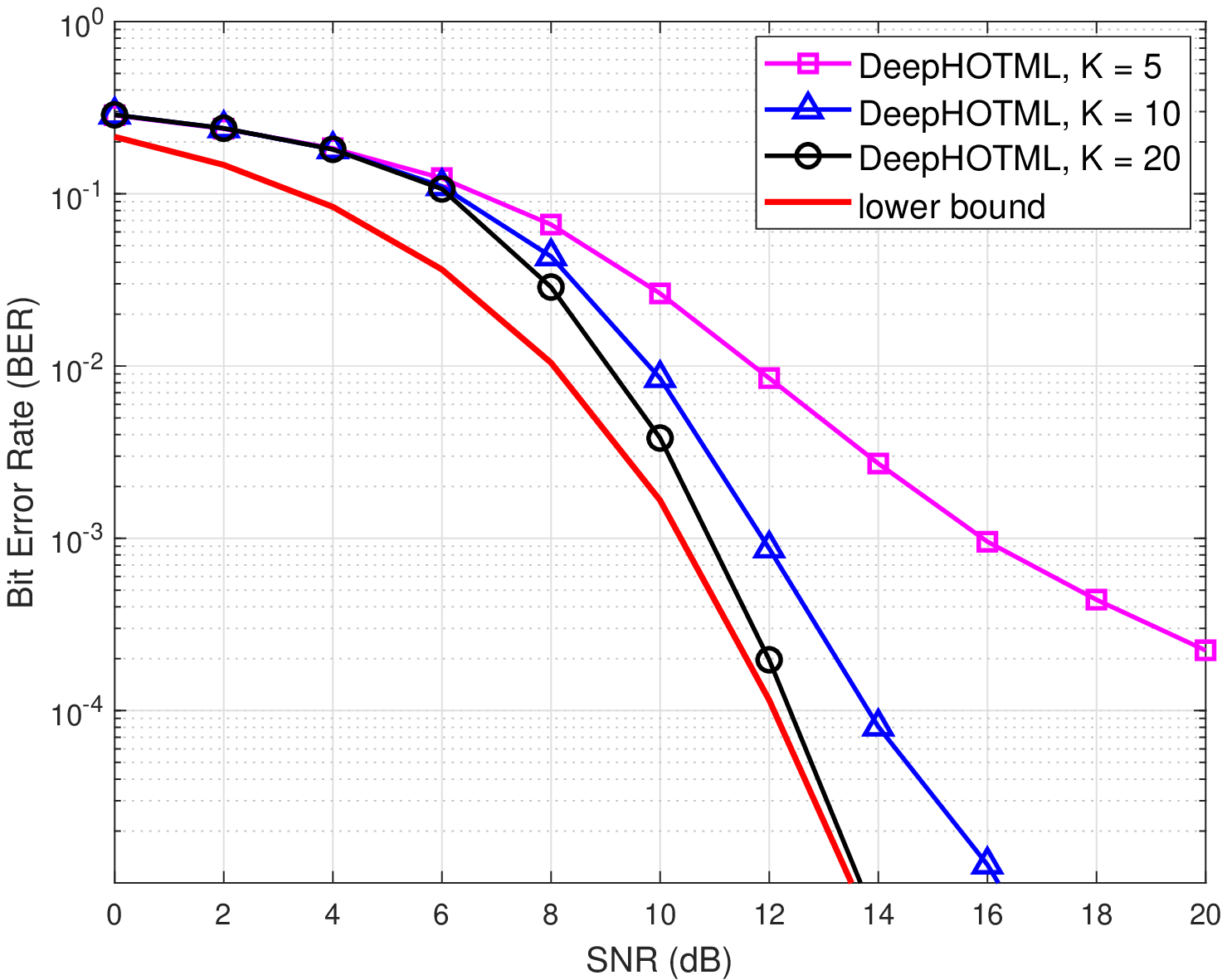}
        \caption{$M= N= 80$}
    \end{subfigure}
    ~ 
    \begin{subfigure}[b]{0.45\textwidth}
        \includegraphics[width=\textwidth]{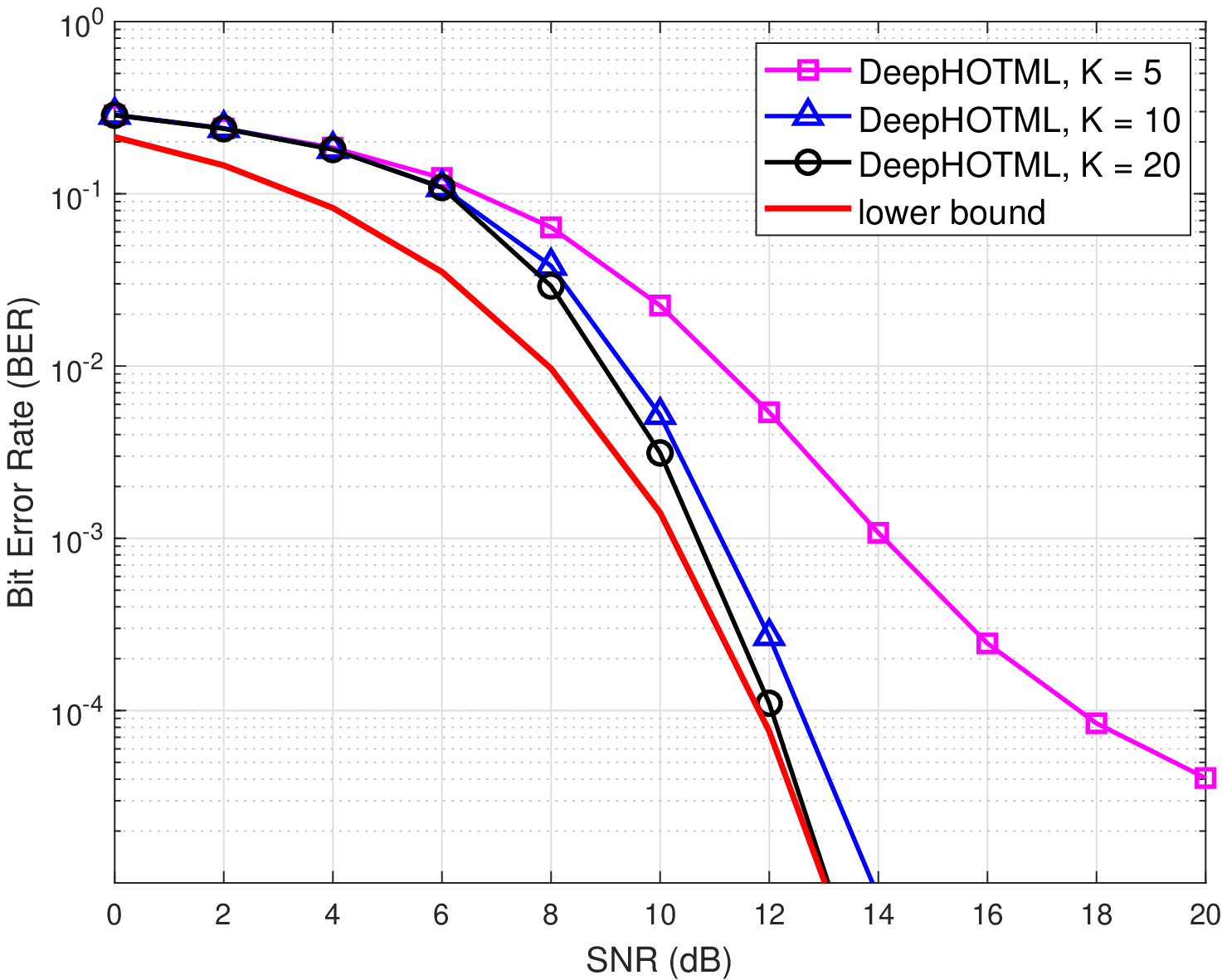}
        \caption{$M= N= 160$}
    \end{subfigure}
    ~ 
    \begin{subfigure}[b]{0.45\textwidth}
        \includegraphics[width=\textwidth]{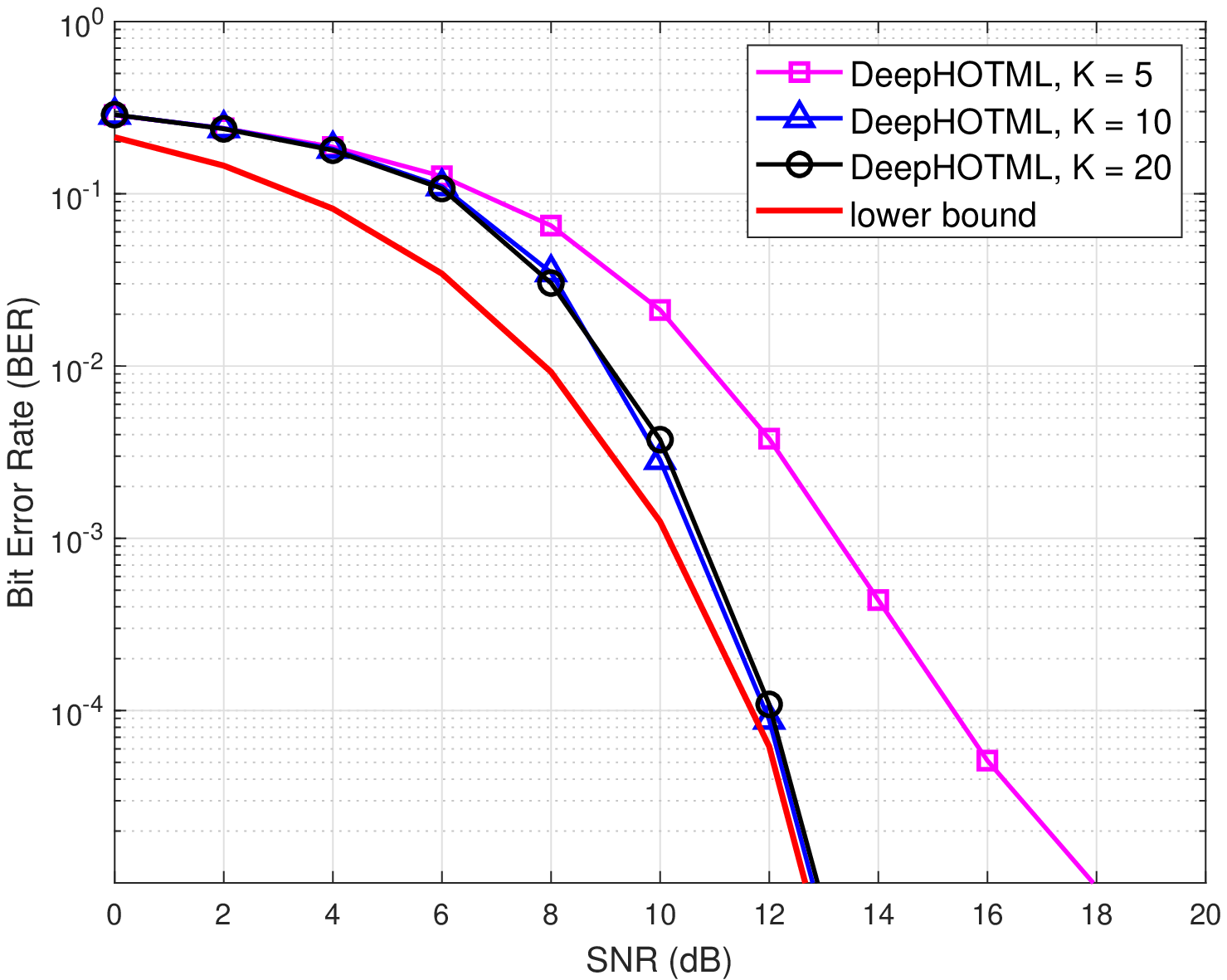}
        \caption{$M= N= 400$}
    \end{subfigure}

    \caption{Classical MIMO detection performance of DeepHOTML under different numbers of layers.}\label{fig:class_ber_deep}
\end{figure}

We finish by showing the BER performance of DeepHOTML for different numbers of layers in Fig.~\ref{fig:class_ber_deep}.
The results indicate that DeepHOTML can achieve promising performance with the number of layers as small as $10$.

\section{Conclusion}

To conclude, we developed a homotopy optimization algorithm for efficient high-performance MIMO detection under one-bit quantized or unquantized observations.
We also studied the possibility of  using deep unfolding to enhance the performance of our homotopy algorithm.
Our numerical results illustrated that the homotopy algorithms, non-deep and deep, achieve promising detection performance.
Also, the deep homotopy algorithm was found to be easy and stable to train for a variety of MIMO settings, and its operational cost very competitive compared to those of many existing MIMO detection methods.
Our present study focused on binary symbol constellations, and we hope that our endeavor would provide further insights into attacking other types of symbol constellations.
In addition it will be interesting to study the extension to the frequency-selective fading scenario, which is much more challenging as revealed in studies such as \cite{studer2016quantized}.

\bibliographystyle{IEEEtran}
\bibliography{refs}

\end{document}